\documentclass[%
reprint,
 amsmath,amssymb,
 aps,
 showpacs
]{revtex4-1}

\usepackage{amsmath}
\usepackage{mathrsfs}
\usepackage{graphicx}
\usepackage{dcolumn}
\usepackage{bm}
\usepackage{ulem}               

\begin{document}

\preprint{APS/123-QED}

\title{
Time-dependent restricted-active-space self-consistent-field theory for laser-driven many-electron dynamics. II. Extended formulation and numerical analysis
}

\author{Haruhide Miyagi}
\author{Lars Bojer Madsen}
\affiliation{Department of Physics and Astronomy, Aarhus University, 8000 Aarhus C, Denmark}

\date{\today}

\begin{abstract}
The time-dependent restricted-active-space self-consistent-field (TD-RASSCF) method is formulated based on the TD variational principle. The SCF based TD orbitals contributing to the expansion of the wave function are classified into three groups, between which orbital excitations are considered with the RAS scheme. In analogy with the configuration-interaction singles (CIS), singles-and-doubles (CISD), singles-doubles-and-triples (CISDT) methods in quantum chemistry, the TD-RASSCF-S, -SD, and -SDT methods are introduced as extensions of the TD-RASSCF doubles (-D) method [Phys. Rev. A {\bf 87}, 062511 (2013)]. Based on an analysis of the numerical cost and test calculations for one-dimensional (1D) models of atomic helium, beryllium, and carbon, it is shown that the TD-RASSCF-S and -D methods are computationally feasible for systems with many electrons and more accurate than the TD Hartree-Fock (TDHF) and TDCIS methods. In addition to the discussion of methodology, an analysis of electron dynamics in the high-order harmonic generation (HHG) process is presented. For the 1D beryllium atom, a state-resolved analysis of the HHG spectrum based on the time-independent HF orbitals shows that while only single-orbital excitations are needed in the region below the cutoff, single- and double-orbital excitations are essential beyond, where accordingly the single-active-electron (SAE) approximation and the TDCIS method break down. On the other hand, the TD-RASSCF-S and -D methods accurately describe the multi-orbital excitation processes throughout the entire region of the HHG spectrum. For the 1D carbon atom, our calculations show that multi-orbital excitations are essential in the HHG process even below the cutoff. Hence, in this test system a very accurate treatment of electron correlation is required. The TD-RASSCF-S and -D approaches meet this demand, while the SAE approximation and the TDCIS method are inadequate.
\end{abstract}


\maketitle

\section{Introduction}

In strong-field and attosecond physics, combinations of femto- and attosecond pulses are used to retrieve time-resolved information about electronic and nuclear motion~\cite{Krausz2009,pascal2012}. Simple physical pictures based on a reduction of the complex many-body dynamics of an atom or molecule under a short pulse to the consideration of a single-active-electron (SAE) have been instrumental in developing the methods of the research field~\cite{Krause1992,Schafer1993,Corkum1993,Lewenstein1994}. Meanwhile, experimental and theoretical investigations are shifting toward more detailed and fundamental analyses of dynamics requiring an account of electron correlation for their complete understanding. For example, attosecond absorption spectroscopy was used to observe and control electronic dynamics~\cite{Chun2013,Ott2013,Chini2013}, the attosecond streak camera was used to determine the intrinsic time delay of photoionization from different orbitals~\cite{Haessler2009,Schultze2010,Neppl2012}, and the mechanism of quantum tunneling was elucidated by the attoclock technique~\cite{Eckle2008,Pfeiffer2012} and by high-order harmonic generation (HHG) spectroscopy~\cite{Shafir2012,Soifer2013,Zhao2013}. These elaborate experiments call for reliable \textit{ab initio} time-dependent (TD) many-electron theories to analyze the results in minute detail without neglecting electron correlation.

Among recent developments of many-electron theories, the TD configuration-interaction singles (TDCIS) method~\cite{Krause2005,Rohringer2006,Gordon2006,Krause2007,Krause2009,Rohringer2009,Greenman2010,Pabst2012,Pabst2012b,Luppi2012,Pabst2013} is the simplest framework beyond the SAE approximation. The TDCIS expansion relative to the Hartree-Fock (HF) ground state provides a set of working equations which are numerically tractable even for large atoms. Another strategy for dealing with the many-electron dynamics is the TD Hartree-Fock (TDHF) method~\cite{Kulander1987,Pindzola1995,Dahlen2001,Zhang2013} or its generalization, i.e., the muticonfigurational TDHF (MCTDHF) method~\cite{Meyer1990,Beck2000,Caillat2005,Meyer2010} where the wave function is expressed by a full-CI expansion with TD orbitals. The key idea in the MCTDHF approach is to optimize the orbitals at each time step by the self-consistent-field (SCF) ansatz, which allows the use of a relatively small number of orbitals for constructing the wave function. While in the perturbative regime for the matter-light interaction, the MCTDHF method can be applied to relatively large systems as illustrated by the computation of inner-shell photoionization cross sections for molecular hydrogen fluoride~\cite{Haxto2012}, in general, due to the full-CI expansion, the number of configurations in the MCTDHF method exponentially increases with respect to the number of electrons, which thus makes the method infeasible for systems having more than a few electrons interacting with intense light fields~\cite{Nest2007,Nest2008,Kato2008,Nest2009,Kato2009a,Sukiasyan2009,Sukiasyan2010,Jhala2010,Hochstuhl2011,Haxto2011,Ulusoy2012,Hochstuhl2014}. As a way to cure the undesirable scaling property of the MCTDHF method, the TD complete-active-space SCF (TD-CASSCF) method was presented~\cite{Sato2013}. In this method the number of configurations is reduced by introducing core orbitals, which, however, eventually will compromise the accurate description of core excitation processes.

Meanwhile, quite a few methodologies have been developed in quantum chemistry to circumvent the expensive full-CI expansion for describing ground-state wave functions of atoms and molecules (see, e.g., the textbook~\cite{Helgaker2000}). The most primitive way is a simple truncation of the CI expansion at a certain excitation level: CI singles (CIS), singles-and-doubles (CISD), singles-doubles-and-triples (CISDT), and so forth. A more advanced concept is the restricted-active-space (RAS) scheme, where the orbitals are classified into three groups, so-called active spaces, between which excitations are allowed with certain restrictions~\cite{Helgaker2000,Hochstuhl2012}. In our recent paper~\cite{Miyagi2013}, as a generalization of the MCTDHF and TD-CASSCF methods, we presented a general formulation of the TD-RASSCF method with special emphasis on the TD-RASSCF-doubles (D) ansatz. In the TD-RASSCF-D method, only double-orbital excitations are allowed between the active spaces, and this restriction reduces the algorithmic complexity associated with solving the working equations of the theory, as will be revisited in this paper. A related method is the orbital adaptive time-dependent coupled-cluster-doubles approach~\cite{Kvaal2012}. Although the orbitals are self-consistently optimized at each time step, due to the lack of the explicit treatment of single-orbital excitations, the numerical accuracy of the doubles-only approaches could be questioned in terms of how reliably the method describes single-electron dynamics which is, e.g., always a key process in strong-field ionization. 

We have two aims with this paper. The first is to present extensions of the TD-RASSCF-D method and to study their numerical performance and accuracy. By including the single-, double-, and triple-orbital excitations, in analogy to the CIS, CISD, and CISDT methods, the TD-RASSCF-S~\cite{Miyagi2014}, -SD, and -SDT methods are formulated based on the Dirac-Frenkel-McLachlan TD variational principle~\cite{Dirac1930, Frenkel1934, McLachlan1964, Lubich2008}. As pointed out in Ref.~\cite{Miyagi2014}, the TD-RASSCF-S method has a special convergence property: the fully converged wave function can be obtained using a small number of orbitals (typically smaller than the number of electrons), and the converged TD-RASSCF-S wave function can be more accurate than the TDCIS wave function in the sense of the TD variational principle. Furthermore, in this paper, TD-RASSCF-D is shown to be more accurate than TD-RASSCF-S in practice, and the TD-RASSCF-S and -D methods are equally feasible for large systems. The more accurate methods, TD-RASSCF-SD, and -SDT, are computationally more expensively but successfully describe phenomena which require a more accurate treatment of the electron correlation. The flexible tunability of the accuracy and cost is a particular advantageous feature of the TD-RASSCF method. The second aim of this paper is to analyze the laser-induced dynamics involved in HHG. By carrying out computations of excitation probabilities and state-resolved analysis of HHG spectra based on field-free time-independent HF orbitals, contributions to the spectra from multi-orbital excitations are clarified. These contributions cause the break-down of the SAE approximation and the TDCIS method. On the other hand, both the TD-RASSCF-S and -D methods succeed in describing the multi-orbital excitations accurately. Another noticeable advantage of the TD-RASSCF method is the gauge independence, i.e., the laser-induced dynamics can be computed without concern about the possible influence of the choice of either length, velocity, or acceleration  gauge.

The paper is organized as follows. The general formulation of the TD-RASSCF method is presented in Sec.~\ref{Overview}. Based on the general concept of the RAS scheme, i.e., assuming no specific restriction for orbital excitations, the equations of motion are derived. Section~\ref{Family} is then devoted to carefully tailoring the equations to each specific RAS scheme, especially focusing on the TD-RASSCF-S, -SD, and -SDT methods. In Sec.~\ref{Numerical performance}, the numerical performances of these methods are examined. To consider the applicability to large systems, the numerical costs are analyzed in Sec.~\ref{Numerical scaling} and detailed in Appendix~\ref{App-SCALE}. For one-dimensional (1D) models of atomic helium, beryllium, and carbon, numerical tests are carried out: computations of the ground-state energies in Sec.~\ref{Ground-state wave function}, calculations of the HHG spectra induced by a laser pulse in Sec.~\ref{High-order harmonic spectrum}. The calculations of laser-induced excitation probabilities and the state-resolved analysis of the HHG spectra are carried out based on time-independent HF orbitals as detailed in Appendix~\ref{App-CISD}. The HHG spectra are also computed within the SAE approximation based on the formulation in Appendix~\ref{App-SAE} as well as with the TDCIS method and compared to the results of the other methods. Based on the TD variational principle, relations among the various TD-RASSCF methods and the TDCIS method are discussed in Sec.~\ref{Accuracy}. The gauge independence of TD-RASSCF and gauge dependence of TDCIS are shown in Appendix~\ref{App-gauge}. Section~\ref{Conclusion} concludes. Atomic units are used throughout unless otherwise stated.

\section{\label{Overview} General formulation}

We start by summarizing the general formulation of the TD-RASSCF theory (see Ref.~\cite{Miyagi2013}). Reviewing the basics provides the necessary background for considering the specific RAS schemes discussed in Sec.~\ref{Family}.

\subsection{\label{EOMformal} Formal derivation of the equations of motion}

\begin{figure}
\begin{center}
\begin{tabular}{c}
\resizebox{70mm}{!}{\includegraphics{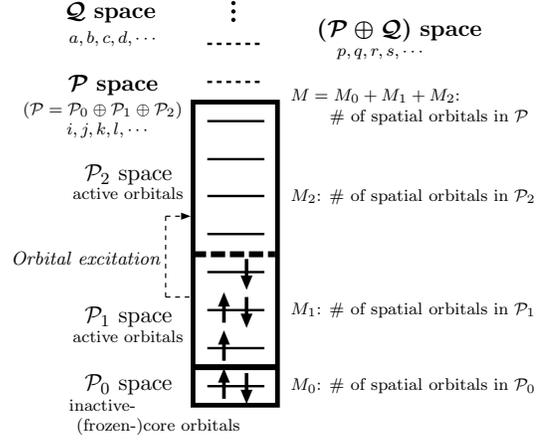}}
\end{tabular}
\caption{
\label{fig_orbital_ras}
Illustration of the division of the single-particle Hilbert space in the TD-RASSCF theory. The wave function is composed of the spin orbitals $|\phi_i(t)\rangle\otimes|\sigma\rangle$ ($i=1,\cdots,M$, and $\sigma=\uparrow,\downarrow$). The $\mathcal{P}$ space spanned by the spatial orbitals consists of three subspaces: an inactive-core space, $\mathcal{P}_0$, and two active spaces, $\mathcal{P}_1$ and $\mathcal{P}_2$, between which orbital transitions are allowed with certain restrictions as specified by the RAS scheme. The rest of the single-particle Hilbert space spanned by the virtual orbitals is referred to as $\mathcal{Q}$ space. The orbitals in either $\mathcal{P}$ or $\mathcal{Q}$ space are labeled $p,q,r,s,\cdots$, while the $\mathcal{P}$-space orbitals are labeled $i,j,k,l,\cdots$, and the $\mathcal{Q}$-space orbitals $a,b,c,d,\cdots$. The numbers of spatial orbitals in the $\mathcal{P}_0$, $\mathcal{P}_1$, and $\mathcal{P}_2$ spaces are expressed by $M_0$, $M_1$ and $M_2$, respectively, and the total number by $M=M_0+M_1+M_2$. The case $M_2=0$ corresponds to having only one active space, i.e., to the TD-CASSCF method, which is further reduced to the MCTDHF method when $M_0=0$ (see Sec.~\ref{P-space orbital eqs for TD-CASSCF}). In this illustration, $(M_0,M_1,M_2)=(1,3,4)$. 
}
\end{center}
\end{figure}

Based on the spin-restricted ansatz, consider the dynamics of an $N_{\rm e}$-electron wave function governed by a TD Hamiltonian, $H(t)$. Using a set of $M(\ge N_{\rm e}/2)$ spatial orbitals, $\big\{|\phi_i(t)\rangle\big\}_{i=1}^M$, the TD-RASSCF wave function is expanded in terms of normalized Slater determinants composed of TD spin-orbitals $|\phi_i(t)\rangle\otimes|\sigma\rangle$ ($i=1,\cdots,M$, and $\sigma=\uparrow,\downarrow$ denoting the spin states),
\begin{eqnarray}
|\Psi(t)\rangle
=
\sum_{{I}\in\mathcal{V}}C_{{I}}(t)|\Phi_{{I}}(t)\rangle,
\label{MCTDHFK_wave_function}
\end{eqnarray}
where the multi-index ${I}$ represents the electronic configurations, and $\mathcal{V}$ is the Fock space spanned by the configurations specified in the considered RAS scheme. The multi-index ${I}$ is decomposed into $\alpha$- and $\beta$-spin strings: ${I}={I}_{\uparrow}\otimes {I}_{\downarrow}$ where ${I}_{\sigma}=(i_1,i_2,\cdots,i_{N_{\sigma}})$ satisfies $1\le i_1<i_2<\cdots<i_{N_{\sigma}}\le M$ and $N_{\uparrow}+N_{\downarrow}=N_{\rm e}$ (see, e.g., Ref.~\cite{Olsen1988}). Following Ref.~\cite{Miyagi2013}, let $\mathcal{P}$ denote the space at time $t$ spanned by $\{|\phi_i(t)\rangle\}_{i=1}^M$ and $\mathcal{Q}$ the rest of the instantaneous single-particle Hilbert space as illustrated in Fig.~\ref{fig_orbital_ras}. The indices $p,q,r,s\cdots$ denote orbitals belonging to either space, while the $\mathcal{P}$-space orbitals are labeled by $i,j,k,l,\cdots$, and the $\mathcal{Q}$-space virtual orbitals by $a,b,c,d,\cdots$. The RAS scheme is defined by dividing the $\mathcal{P}$ space into three subspaces: the inactive-core space, $\mathcal{P}_0$, and the two active spaces, $\mathcal{P}_1$ and $\mathcal{P}_2$, between which orbital excitations are allowed subject to the restrictions specified by the RAS scheme. If the core electrons do not influence the dynamics, the frozen-core approximation, i.e., the use of the time-independent HF orbitals as the $\mathcal{P}_0$-space orbitals would be a viable route to reduce the computational complexity~\cite{Sato2013}. The numbers of spatial orbitals in $\mathcal{P}_0$, $\mathcal{P}_1$, and $\mathcal{P}_2$ are denoted by $M_0$, $M_1$, and $M_2$, respectively (hence, $M=M_0+M_1+M_2$). The case with one active space and a core, i.e., the TD-CASSCF method, corresponds in our formalism to $M_2=0$ and $M_0+M_1=M$. The case with just one active space and no core, i.e., the MCTDHF method, corresponds to $M_0=M_2=0$ and $M_1=M$. 

The set of equations obeyed by the CI-expansion coefficients and the orbitals in Eq.~\eqref{MCTDHFK_wave_function} are derived based on the Dirac-Frenkel-McLachlan TD variational principle \cite{Dirac1930, Frenkel1934, McLachlan1964, Lubich2008}. For brevity, henceforth, explicit time-dependence of parameters and operators is dropped as long as it causes no confusion. We define an action functional
\begin{eqnarray}
&&
\hspace{-4mm}
\mathcal{S}\big[\{C_{{I}}\},\{\phi_i\},\{\varepsilon^i_j\}\big]=
\nonumber \\
&&
\hspace{0mm}
\int_0^T
\Bigg[
\langle\Psi|\Bigg( i\frac{\partial}{\partial t}-H\Bigg)
|\Psi\rangle
+\sum_{ij} \varepsilon^i_j\Big(
\langle\phi_i|\phi_j\rangle
-\delta^i_j
\Big)
\Bigg]dt,
\label{functional_CI}
\end{eqnarray}
where $\varepsilon^i_j$ is a Lagrange multiplier which ensures orthonormality among the $\mathcal{P}$-space orbitals during the time interval $[0,T]$. The stationary conditions, $\delta\mathcal{S}/\delta C_{{I}}^*=0$ and $\delta\mathcal{S}/\langle\delta\phi_i|=0$, respectively, result in the amplitude equations 
\begin{eqnarray}
i\dot{C}_{{I}}
+\langle\Phi_{{I}}|(iD-H)|\Psi\rangle=0,
\label{amp_eqs1}
\end{eqnarray}
and the orbital equations
\begin{eqnarray}
\sum_q
|\phi_q\rangle
\langle\Psi_i^q|
\Bigg(
i\sum_{{I}\in\mathcal{V}}\dot{C}_I|\Phi_{{I}}\rangle
&&+(iD-H)|\Psi\rangle
\Bigg)
\nonumber \\
&&
\hspace{0mm}
+\sum_{j} |\phi_j\rangle \varepsilon_j^i=0,
\label{orb1}
\end{eqnarray}
where $\langle\Psi_i^q|\equiv\langle\Psi|E^q_i$ is the \textit{one-particle-one-hole state}, and $E^q_i$ is the spin-free excitation operator defined by
\begin{eqnarray}
E_p^q&=&\sum_{\sigma=\uparrow,\downarrow}c_{p\sigma}^{\dagger}c_{q\sigma}, 
\label{spin-free1}
\end{eqnarray}
with $c_{p\sigma}$ ($c^{\dagger}_{p\sigma}$) being the annihilation (creation) operator of an electron in the spin orbital $|\phi_p(t)\rangle\otimes|\sigma\rangle$. Both Eqs.~\eqref{amp_eqs1} and \eqref{orb1} contain the orbital-time-derivative operator,
\begin{eqnarray}
D
=\sum_{pq}\eta_q^p E_p^q
\label{definition_D}
\end{eqnarray}
with $\eta^p_q\big(=-(\eta^q_p)^*\big)=\langle\phi_p|\dot{\phi}_q\rangle$. The orbital equations (\ref{orb1}) consist of two parts: the $\mathcal{Q}$-space orbital equations 
\begin{eqnarray}
\langle\Psi_i^a|(iD-H)|\Psi\rangle
=0,
\label{Q_orb_eqs0}
\end{eqnarray}
and the $\mathcal{P}$-space orbital equations
\begin{eqnarray}
\langle\Psi|(iD-H)|\Psi_{j''}^{i'}\rangle
-
\langle\Psi^{j''}_{i'}|(iD-H)|\Psi\rangle
=
i\dot{\rho}^{j''}_{i'}
\label{orb_eqs_P}
\end{eqnarray}
with the time-derivative of the density matrix
\begin{eqnarray}
\dot{\rho}^{j''}_{i'}=
\sum_{{I}\in\mathcal{V}}
\left(
\dot{C}^*_{{I}}
\langle\Phi_{{I}}|\Psi^{i'}_{j''}\rangle
+
\langle\Psi^{j''}_{i'}|\Phi_{{I}}\rangle\dot{C}_{{I}}
\right).
\label{time-derivative-density}
\end{eqnarray}
In Eqs.~\eqref{orb_eqs_P} and \eqref{time-derivative-density}, the orbitals labeled by single and double primed index $i'$ and $j''$ belong to different subspaces, otherwise the expression \eqref{orb_eqs_P} gives an identity, not an equation (see Sec.~\ref{Family}).

Before closing this section, it should be noted that, whereas the $\mathcal{Q}$-space orbital equations \eqref{Q_orb_eqs0} are coupled only within each other, the amplitude and the $\mathcal{P}$-space orbital equations, i.e., Eqs.~\eqref{amp_eqs1} and \eqref{orb_eqs_P}, compose a set of coupled equations linked together via $\dot{\rho}^{j''}_{i'}$. As discussed in Sec.~\ref{Family}, it is anything but trivial to eliminate this latter coupling except in a few specific RAS schemes.

\subsection{\label{Family} Explicit formula}

For practical implementation of the method, we need explicit expressions for the equations of motion. For completeness, these are given below. The details of the derivation are given in Ref.~\cite{Miyagi2013}. We consider a specific Hamiltonian composed by one- and two-body operators expressed in second quantization as
\begin{eqnarray}
H(t)
=
\sum_{pq} h^p_q(t) E_p^q
+
\frac{1}{2}
\sum_{pqrs} v^{pr}_{qs}(t) E_{pr}^{qs},
\label{Hamiltonian1}
\end{eqnarray}
where the spin-free excitation operators are defined by Eq.~\eqref{spin-free1} and Eq.~\eqref{spin-free2}:
\begin{eqnarray}
E_{pr}^{qs}&=&\sum_{\sigma=\uparrow,\downarrow}\sum_{\tau=\uparrow,\downarrow}
c_{p\sigma}^{\dagger}c_{r\tau}^{\dagger}c_{s\tau}c_{q\sigma}.
\label{spin-free2}
\end{eqnarray}
The matrix elements in Eq.~\eqref{Hamiltonian1} are given by $h^{p}_{q}(t)=\int\phi_p^*({\bm r},t)h({\bm r},t)\phi_q({\bm r},t)d{\bm r}$, and $v^{pr}_{qs}(t)=\iint\phi_p^*({\bm r}_1,t)\phi_r^*({\bm r}_2,t)v({\bm r}_1,{\bm r}_2)\phi_q({\bm r}_1,t)\phi_s({\bm r}_2,t) d{\bm r}_1d{\bm r}_2$. 

Substituting Eqs.~\eqref{definition_D} and \eqref{Hamiltonian1} into Eq.~\eqref{amp_eqs1}, we obtain the explicit expression for the amplitude equations
\begin{eqnarray}
i\dot{C}_{{I}}
=
\sum_{ij} (h^i_j -i\eta_j^i) \langle\Phi_{{I}}|E_i^j|\Psi\rangle
+
\frac{1}{2}
\sum_{ijkl} v^{ik}_{jl} \langle\Phi_{{I}}|E_{ik}^{jl}|\Psi\rangle.
\nonumber \\
\label{amp_orb}
\end{eqnarray}
Likewise by using Eqs.~\eqref{definition_D} and \eqref{Hamiltonian1} and defining the projection operator $Q=1-\sum_{i} |\phi_i\rangle\langle\phi_i|$, the $\mathcal{Q}$-space orbital equations \eqref{Q_orb_eqs0} lead to
\begin{eqnarray}
i\sum_j Q|\dot{\phi}_j\rangle\rho^j_i
=\sum_jQh(t)|\phi_j\rangle\rho^j_i
+\sum_{jkl}QW^k_l|\phi_j\rangle
\rho^{jl}_{ik}
\nonumber \\
\label{Q_orbital_eq1}
\end{eqnarray}
with the density matrices $\rho^j_i\equiv\langle\Psi|E^j_i|\Psi\rangle$ and $\rho^{jl}_{ik}\equiv\langle\Psi|E^{jl}_{ik}|\Psi\rangle$, and the mean-field operator $W^k_l({\bm r},t)=\int\phi_k^*({\bm r}',t)v({\bm r},{\bm r}')\phi_l({\bm r}',t) d{\bm r}'$. Finally, we express the $\mathcal{P}$-space orbital equations \eqref{orb_eqs_P} as
\begin{eqnarray}
\sum_{k''l'} (h^{k''}_{l'}-i\eta^{k''}_{l'}) A^{l'j''}_{k''i'}
+\sum_{klm}
(v^{j''m}_{kl} \rho^{kl}_{i'm} -v^{kl}_{i'm} \rho^{j''m}_{kl})
=i\dot{\rho}_{i'}^{j''}
\nonumber \\
\label{P_orb_eqs5}
\end{eqnarray}
with 
$
A^{l'j''}_{k''i'}
=
\langle\Psi|\big[E^{j''}_{i'},E^{l'}_{k''}\big]|\Psi\rangle
=
\delta^{j''}_{k''}\rho^{l'}_{i'}-\delta^{l'}_{i'}\rho^{j''}_{k''}
$. After solving the $\mathcal{Q}$- and $\mathcal{P}$-space orbital equations [Eqs.~\eqref{Q_orbital_eq1} and \eqref{P_orb_eqs5}], the time-derivatives of the $\mathcal{P}$-space orbitals are obtained as
\begin{eqnarray}
|\dot{\phi}_i\rangle
=(P+Q)|\dot{\phi}_i\rangle 
=
\sum_j |\phi_j\rangle \eta^j_i
+Q|\dot{\phi}_i\rangle.
\label{total_0}
\end{eqnarray}

\section{\label{Family} Specific RAS schemes}

\begin{figure*}[t]
\begin{center}
\begin{tabular}{c}
\resizebox{130mm}{!}{\includegraphics{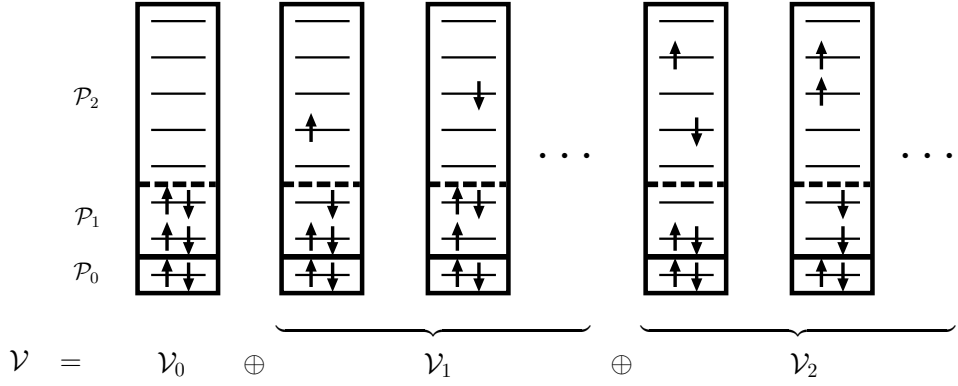}}
\end{tabular}
\caption{
\label{fig_ras_directsum}
Decomposition of the Fock space. This example shows the case of the TD-RASSCF-SD method, i.e., the TD-RASSCF$(N=2)$ method, with $(M_0,M_1,M_2)=(1,2,5)$ for a six-electron system. The Fock space is expressed by a direct sum, $\mathcal{V}=\mathcal{V}_0\oplus\mathcal{V}_1\oplus\mathcal{V}_2$, where $\mathcal{V}_n$ ($n=0,1,$ and $2$) means a subspace consisting of all possible $n$-fold-orbital excited configurations.
}
\end{center}
\end{figure*}

As discussed in Sec.~\ref{Overview}, the presence of $\dot{\rho}^{j''}_{i'}$ couples the amplitude and the $\mathcal{P}$-space orbital equations [see either Eqs.~\eqref{amp_eqs1} and \eqref{orb_eqs_P} or Eqs.~\eqref{amp_orb} and \eqref{P_orb_eqs5}], 
which requires elaborate implicit integration schemes. The difficultly can be resolved if the coupling can be removed. Substituting Eq.~\eqref{time-derivative-density} into Eq.~\eqref{orb_eqs_P}, and using Eq.~\eqref{amp_eqs1}, we arrive at the expression 
\begin{eqnarray}
&&\langle\Psi|(iD-H)(1-\Pi)|\Psi^{i'}_{j''}\rangle
\nonumber \\
&&
\hspace{15mm}
-
\langle\Psi_{i'}^{j''}|(1-\Pi)(iD-H)|\Psi\rangle
=0
\label{P_orb_eqs03}
\end{eqnarray}
with the projection operator $\Pi$ defined by
\begin{eqnarray}
\Pi=
\sum_{{I}\in\mathcal{V}}|\Phi_{{I}}\rangle\langle\Phi_{{I}}|.
\label{Pi}
\end{eqnarray}
Note that Eq.~\eqref{P_orb_eqs03} becomes an identity if $|\phi_{i'}\rangle$ and $|\phi_{j''}\rangle$ belong to the same subspace. Equation~\eqref{P_orb_eqs03} is formally free from the coupling except the presence of $\Pi$, whose operation is specifically defined in each considered RAS scheme. We will see that, in particular RAS schemes, Eq.~\eqref{P_orb_eqs03} can be rewritten in explicit forms which are appropriate for numerical treatments.

\subsection{\label{P-space orbital eqs for TD-CASSCF} TD-CASSCF and MCTDHF methods}

Both the TD-CASSCF method and MTDHF approaches are obtained as special cases of the TD-RASSCF method. The TD-CASSCF method is obtained by setting $M_0+M_1=M$ and $M_2=0$ (see Fig.~\ref{fig_orbital_ras}). Because there is no orbital transitions between $\mathcal{P}_0$ and $\mathcal{P}_1$ spaces, $\rho^{j''}_{i'}$ and $\dot{\rho}^{j''}_{i'}$ are absent~\cite{Miyagi2013,Sato2013}, and the $\mathcal{P}$-space orbital equations \eqref{P_orb_eqs5} simplify to
\begin{eqnarray}
\sum_{k''l'} (h^{k''}_{l'}-i\eta^{k''}_{l'}) A^{l'j''}_{k''i'}
+\sum_{klm}
(v^{j''m}_{kl} \rho^{kl}_{i'm} -v^{kl}_{i'm} \rho^{j''m}_{kl})
=0.
\nonumber \\
\label{P_orb_eqs6}
\end{eqnarray}
If there is no core, i.e., in the MCTDHF method with $M_1=M$ and $M_0=M_2=0$, Eqs.~\eqref{orb_eqs_P}, \eqref{P_orb_eqs5}, and \eqref{P_orb_eqs03} become identities. The $\mathcal{P}$-space orbital equations hence disappear, and the values of $\eta_i^j$ are thus often chosen as zeros \cite{Beck2000, Meyer2010}. In summary, the TD-CASSCF method is expressed by a set of equations of motion, Eqs.~\eqref{amp_orb}, \eqref{Q_orbital_eq1}, and \eqref{P_orb_eqs6}, and the MCTDHF method by Eqs.~\eqref{amp_orb} and \eqref{Q_orbital_eq1}. 

\subsection{\label{P-space orbital eqs for TD-RASSCF-D} TD-RASSCF-D method}

The TD-RASSCF-D method~\cite{Miyagi2013} takes into account all possible double-orbital excitations from $\mathcal{P}_1$ to $\mathcal{P}_2$. Because the occupation number in the $\mathcal{P}_2$ space is zero or two, the matrix elements $\rho^{j''}_{i'}$ and $\dot{\rho}^{j''}_{i'}$ disappear, which results in the $\mathcal{P}$-space orbital equations exactly of the form expressed by Eq.~\eqref{P_orb_eqs6}. The TD-RASSCF-D method thus requires one to solve the set of equations of motion given by Eqs.~\eqref{amp_orb}, \eqref{Q_orbital_eq1}, and \eqref{P_orb_eqs6}.

\subsection{\label{P-space orbital eqs for  TD-RASSCF-S, -SD, and -SDT} TD-RASSCF-S, -SD, and -SDT methods}

Consider a series of methods, TD-RASSCF$(N)$ ($N=1,2,\cdots,\min\{M_1,M_2\}$), defined by prohibiting $K$-fold-orbital excitations ($K>N$). In analogy to the CIS, CISD, and CISDT methods (see, e.g., Ref.~\cite{Helgaker2000}), the cases of $N=1,2$, and $3$ are specifically denoted TD-RASSCF-S, -SD, and -SDT methods, respectively. In the presence of a core ($M_0\ne0$), the $\mathcal{P}$-space orbital equations result in Eq.~\eqref{P_orb_eqs6} for the set of indices $(i',j'')$ with either $|\phi_{i'}(t)\rangle$ or $|\phi_{j''}(t)\rangle$ belonging to $\mathcal{P}_0$. When neither $|\phi_{i'}(t)\rangle$ nor $|\phi_{j''}(t)\rangle$ belongs to $\mathcal{P}_0$, special treatments are needed in the TD-RASSCF$(N)$ method. In this case, the Fock space $\mathcal{V}$ is decomposed into a direct sum of $N+1$ subspaces (see Fig.~\ref{fig_ras_directsum}):
\begin{eqnarray}
\mathcal{V}=
\mathcal{V}_0\oplus
\mathcal{V}_1\oplus \cdots\oplus
\mathcal{V}_N,
\label{direct_sum_K}
\end{eqnarray}
where $\mathcal{V}_n$ ($n=0, 1, \cdots, N$) denotes a subspace spanned by all possible configurations with $n$-fold-orbital excitations from $\mathcal{P}_{1}$ to $\mathcal{P}_{2}$. In Eq.~(\ref{P_orb_eqs03}), for $|\phi_{i'}(t)\rangle\in\mathcal{P}_{1}\wedge|\phi_{j''}(t)\rangle\in\mathcal{P}_{2}$, $\langle\Psi_{i'}^{j''}|$ contains $(N+1)$-fold-orbital excited configurations composing $\mathcal{V}_{N+1}(\not\subset\mathcal{V})$, but $|\Psi_{j''}^{i'}\rangle$ does not, hence $\langle\Psi_{i'}^{j''}|(1-\Pi)\ne0$ and $(1-\Pi)|\Psi_{j''}^{i'}\rangle=0$. Equation~\eqref{P_orb_eqs03} therefore simplifies to
\begin{eqnarray}
\langle \Psi_{i'}^{j''}|(1-\Pi)(iD-H)|\Psi\rangle
=0,
\label{MCTDHF(n)-P-formal}
\end{eqnarray}
with 
\begin{eqnarray}
\langle\Psi_{i'}^{j''}|(1-\Pi)
=
\sum_{{I}\in \mathcal{V}_{N}}C_{{I}}^* \langle\Phi_I|E_{i'}^{j''},
\label{PiPsi}
\end{eqnarray}
where the summation in Eq.~\eqref{PiPsi} runs within the $N$-fold-orbital excited configurations. Substituting Eqs.~\eqref{definition_D} and \eqref{Hamiltonian1} into Eq.~\eqref{MCTDHF(n)-P-formal}, we obtain the explicit form
\begin{eqnarray}
\sum_{k''l'} 
(i\eta^{k''}_{l'}-h^{k''}_{l'}) \zeta^{l'j''}_{k''i'}
=
\frac{1}{2}
\sum_{klmn}
v_{ln}^{km}  
\zeta_{kmi'}^{lnj''},
\label{MCTDHF(n)-P-formal2}
\end{eqnarray}
with the fourth- and sixth-order tensors defined by:
\begin{eqnarray}
\zeta^{l'j''}_{k''i'}&=&\langle\Psi_{i'}^{j''}|(1-\Pi)E_{k''}^{l'}|\Psi\rangle, \label{4th-tensor} \\
\zeta_{kmi'}^{lnj''}&=& \langle\Psi_{i'}^{j''}|(1-\Pi)E^{ln}_{km}|\Psi\rangle. \label{6th-tensor}
\end{eqnarray}
In Eq.~\eqref{6th-tensor} $E^{ln}_{km}$ should excite one or two orbitals in the ket-vector $|\Psi\rangle$ from $\mathcal{P}_1$ to $\mathcal{P}_2$, because $\zeta_{kmi'}^{lnj''}$ is zero otherwise. Exploiting this fact and also the antisymmetry property for exchanging indices, the right hand side of Eq.~(\ref{MCTDHF(n)-P-formal2}) is rewritten as
\begin{eqnarray}
\frac{1}{2}
\sum_{klmn}
v_{ln}^{km}  
\zeta_{kmi'}^{lnj''}
=
\sum_{\substack{k'm'' \\ l'<n'}}
\big(v_{l'n'}^{k'm''} -v_{n'l'}^{k'm''}\big)\zeta_{k'm''i'}^{l'n'j''}
\nonumber \\
+
\sum_{\substack{k''<m'' \\ l'n''}}
\big(v_{l'n''}^{k''m''} -v_{n''l'}^{k''m''}\big)\zeta_{k''m''i'}^{l'n''j''}.
\label{P-space orbital equations}
\end{eqnarray}
Instead of  directly preparing all values of $\zeta_{kmi'}^{lnj''}$ and then computing the right hand side of Eq.~\eqref{MCTDHF(n)-P-formal2}, one should use Eq.~\eqref{P-space orbital equations} to reduce the numerical cost.

\section{\label{Numerical performance} Numerical performance}

\subsection{\label{Numerical scaling} Analysis of computational cost}

Before starting numerical implementations of the TD-RASSCF methods, it is important to know which part of the equations is the most time consuming to evaluate and to estimate how the computational cost depends on the considered RAS scheme. For the analysis, we consider a 1D model of an $N_{\rm e}$-electron atom, and each orbital is expanded in a discrete-variable-representation (DVR) basis set~\cite{Light1985} with $N_{\rm DVR}$ quadrature points. To obtain converged results for laser-induced dynamics, typically $N_{\rm DVR}=O(10^3)$ and the number of spatial orbitals will be at least $M= N_{\rm e}$ or more. For simplicity, we suppose closed-shell systems ($N_{\rm e}$ is even). We consider a simple partitioning with $(M_0,M_1,M_2)=(0,N_{\rm e}/2,M-N_{\rm e}/2)$ for the TD-RASSCF-S, -D, -SD, -SDT, $\cdots$, and $(M_0,M_1,M_2)=(0,M,0)$ for the MCTDHF method.

To solve the equations of motion in the TD-RASSCF-D and MCTDHF methods, the number of operations at each time step is estimated as follows (see Appendix~\ref{App-SCALE}): Preparation of the values of $v_{jl}^{ik}$ and then solving the amplitude equations~\eqref{amp_orb} require $M^4N_{\rm DVR}^2$ and $M^4\dim \mathcal{V}$ operations, respectively. Similarly, the preparation of the values of $\rho^{jl}_{ik}$ and the integration of the $\mathcal{Q}$-space orbital equations~\eqref{Q_orbital_eq1} need $M^4\dim \mathcal{V}$ and $M^4N_{\rm DVR}^2$ operations, respectively. The cost for integrating the $\mathcal{P}$-space orbital equations \eqref{P_orb_eqs6} is negligible compared to the operations above. In both methods, the total overhead is therefore approximately $2M^4(N_{\rm DVR}^2+\dim \mathcal{V})$. In the MCTDHF method, with increasing $N_{\rm e}$, the number of configurations increases exponentially, $\dim \mathcal{V}=O(M^{N_{\rm e}})$, which makes the application difficult to large systems. The TD-RASSCF-D method cures this undesirable scaling property and gives $\dim \mathcal{V}=O(N_{\rm e}^2M^2)$. In the TDHF method, since there is no amplitude equation, the total number of operations is about $(N_{\rm e}/2)^4N_{\rm DVR}^2$ for integrating the $\mathcal{Q}$-space orbital equations.

In the TD-RASSCF-S, -SD, and -SDT methods, updating the values of the sixth-order tensors, $\zeta_{kmi'}^{lnj''}$ [Eq.~\eqref{6th-tensor}], requires about $M^6\dim \mathcal{V}_N$ operations ($N=1,2,$ and $3$ for the TD-RASSCF-S, -SD, and -SDT methods, respectively). Noting that $\dim \mathcal{V}_N=O(N_{\rm e}^NM^N)$, it is seen that the computation of $\zeta_{kmi'}^{lnj''}$ is the most demanding step. That is, although the total number of the single-orbital excitations is small, $\dim\mathcal{V}_1=O(N_{\rm e}M)$, their inclusion generates a large computational cost due to the emergence of the sixth-order tensor. 

\begin{figure}
\begin{center}
\begin{tabular}{c}
\resizebox{85mm}{!}{\includegraphics{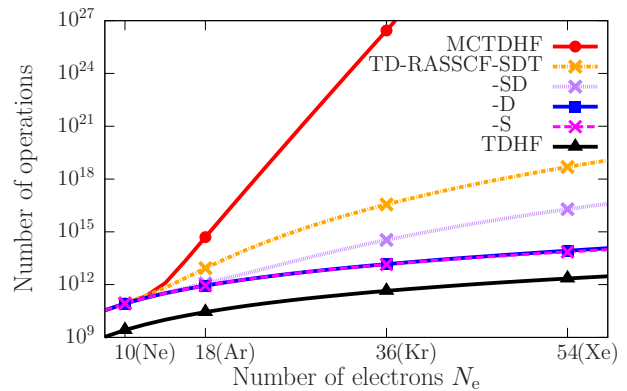}}
\end{tabular}
\caption{
\label{fig_scale}
(Color online) The number of operations at each time step for solving the equations of motion as a function of the number of electrons [Eqs.~\eqref{dimV_HF}--\eqref{dimV_FCI}] for the 1D realizations of the methods considered in this work. To produce the plot, $N_{\rm DVR}=2048$ is fixed. The number of orbitals is set to be $N_{\rm e}/2$ for the TDHF method and $M=N_{\rm e}$ for the other methods. In the TD-RASSCF-S, -D, -SD, -SDT methods, the $\mathcal{P}$ space was partitioned as $(M_0,M_1,M_2)=(0,N_{\rm e}/2,N_{\rm e}/2)$. Data points are shown only for rare gas atoms ($N_{\rm e}=10,18,36$, and $54$).
}
\end{center}
\end{figure}

The numbers of operations required at each time step are summarized in Appendix~\ref{App-SCALE} for every method [Eqs.~\eqref{dimV_HF}--\eqref{dimV_FCI}] and plotted in Fig.~\ref{fig_scale} as a function of the number of electrons with $N_{\rm DVR}=2048$. The number of orbitals is $N_{\rm e}/2$ in the TDHF method and is set to $M=N_{\rm e}$ in the other methods. Note that, for closed-shell systems, the TD-RASSCF-S wave function is fully converged with $M=N_{\rm e}$ as shown in Ref.~\cite{Miyagi2014} and as will be revisited in Sec.~\ref{Ground-state wave function}. Although a closed-shell can be realised for every even $N_{\rm e}$ in the 1D model, it can be formed only in rare gas atoms in the realistic 3D cases. Hence the data points in Fig.~\ref{fig_scale} are shown only for $N_{\rm e}=10,18,36$, and $54$. The TDHF method has already been successfully used for computing strong-field ionization of CO molecules ($N_{\rm e}=14$)~\cite{Zhang2013}, and the scaling in Fig.~\ref{fig_scale} might indicate the potential applicability of the TD-RASSCF-S and -D methods to such or even larger realistic 3D systems. The TD-RASSCF-SD method can possibly be applied as well. Remember that, except in the TD-RASSCF-S method, the use of more orbitals ($M>N_{\rm e}$) gives more accurate results but is computationally more expensive, so that the use of inactive-core or frozen-core orbitals will be indispensable in such computations for very large systems.

To expand the orbitals, it is possible to use any kind of basis set. A change of basis will
change the numerical cost. Consider the evaluation of $v_{jl}^{ik}$. The use of the DVR basis functions associated with the Legendre polynomials, for instance, requires $N_{\rm DVR}^2$ operations to evaluate the value because the DVR is diagonal in the position representation~\cite{Light1985}. On the other hand, the use of the Legendre polynomials as a basis set requires $N_{\rm DVR}^4$ operations. To expand the orbitals in a large box, the best choice will thus be the DVR or the more sophisticate finite-element DVR basis functions~\cite{Hochstuhl2011,Haxto2011}. Finally note that, in computations for realistic 3D Coulomb systems, two additional degrees of freedom need to be taken into account in each orbital. For example, in terms of the angular momentum representation. In this case, therefore, the computations of $W^k_l({\bm r})$ and $v^{ik}_{jl}$ will be another bottle-neck. Instead of carrying out direct integration, one should, in this case, consider the Poisson equation, $\Delta W^k_l({\bm r})=-4\pi \phi_k^*({\bm r})\phi_l({\bm r})$, which gives simple expressions to $W^k_l({\bm r})$ and $v^{ik}_{jl}$ in the DVR \cite{McCurdy2004}.

\subsection{\label{Ground-state wave function} Ground-state energy}

\begin{table*}[t]
\caption{\label{table_Be}
Ground-state energy (in atomic units) of the 1D beryllium atom ($Z=N_{\rm e}=4$). The integers in parentheses below each energy show the number of configurations. In the TD-RASSCF-S, -D, -SD, and -SDT calculations, the partition is set to be $(M_0,M_1,M_2)=(0,2(=N_{\rm e}/2),M-2)$. When $M=2$, the methods reduce to the TDHF method, which gives the HF ground-state energy $-6.739450$. The TD-RASSCF-S method gives a converged result for $M_2\ge M_1$ as indicated by the underlined energies. When $M=N_{\rm e}/2+1=3$, note the following facts: (i) TD-RASSCF-SD and MCTDHF are the same method, (ii) TD-RASSCF-D and MCTDHF (TD-RASSCF-SD) are different methods but theoretically equivalent (see Ref.~\cite{Miyagi2013}) and hence give the same energy value as indicated by symbols $\flat$ (see text), and (iii) TD-RASSCF-SDT can not be defined (where the table thus remains blank). 
}
\begin{ruledtabular}
\begin{tabular}{rlccccccc}
&&\multicolumn{6}{c}{$M$}
\\[0pt]
\cline{3-8}
\\[-7pt]
Method & \hspace{-4mm}$(M_0,M_1,M_2)$
\hspace{0mm}	& 3			& 4				& 8			 	& 12			& 16				& 20 \\ \\[-7pt] \hline \\[-7pt]
TD-RASSCF-S & \hspace{-4mm}$(0,2,M-2)$ & $-6.771254$	& $\underline{-6.773288}$		& $\underline{-6.773288}$		& $\underline{-6.773288}$	& $\underline{-6.773288}$		& $\underline{-6.773288}$	\\
			 && $(5)$	& $(9)$		& $(25)$		& $(41)$ 	 & $(57)$		& $(73)$	\\
-D& \hspace{-4mm}$(0,2,M-2)$ & $-6.771296^{\flat}$	& $-6.779805$		& $-6.784501$		& $-6.784533$	& $-6.784534$	& $-6.784534$  \\
			&& $(5)$	& $(19)$		& $(175)$	& $(491)$ 	& $(967)$	& $(1603)$ \\
-SD& \hspace{-4mm}$(0,2,M-2)$ & $-6.771296^{\flat}$			& $-6.780026$		& $-6.784667$		& $-6.784697$	& $-6.784698$	& $-6.784698$  \\
			&& $(9)$			& $(27)$		& $(199)$	& $(531)$& $(1023)$	& $(1675)$ 	\\
-SDT& \hspace{-4mm}$(0,2,M-2)$ & $$			& $-6.780026$		& $-6.785038$		& $-6.785074$	& $-6.785074$		& $-6.785075$ 	\\
			&& $$			& $(35)$		& $(559)$	& $(2331)$& $(6119)$	& $(12691)$	\\
MCTDHF &\hspace{-4mm}$(0,M,0)$	& $-6.771296^{\flat}$	& $-6.780026$		& $-6.785041$		& $-6.785077$	& $-6.785078$	& $-6.785078$	\\
			&& $(9)$	& $(36)$	& $(784)$	& $(4356)$	& $(14400)$	& $(36100)$	\\ \\[-7pt]
			\end{tabular}
\end{ruledtabular}
\end{table*}

\begin{table*}[t]
\caption{\label{table_C}
Ground-state energy (in atomic units) of the 1D carbon atom ($Z=N_{\rm e}=6$). The integers in parentheses below each energy show the number of configurations. In the TD-RASSCF-S, -D, -SD, and -SDT calculations, the partition $(M_0,M_1,M_2)$ satisfies a condition $M_0+M_1=N_{\rm e}/2=3$. The upper and lower parts consist of the results for $M_0=0$ and $1$, respectively. When $M=3$, the methods reduce to the TDHF theory, which gives the HF ground-state energy $-13.23117$. The TD-RASSCF-S method gives a converged result for $M_2\ge M_1$ as indicated by the underlined energies. When $M=N_{\rm e}/2+1=4$, note the following facts: (i) TD-RASSCF-SD and MCTDHF are the same method, (ii) TD-RASSCF-D and MCTDHF (TD-RASSCF-SD) are different methods but theoretically equivalent (see Ref.~\cite{Miyagi2013}) and hence give the same energy value as indicated by symbols $\flat$, (iii) TD-RASSCF-SDT can not be defined (where the table thus remains blank), and (iv) TD-RASSCF-S gives the same energy value as marked by symbols $\sharp$ irrespective of the value of $M_0(< N_{\rm e}/2)$. The TD-RASSCF-STD calculation with $(M_0,M_1,M_2)=(0,3,2)$ is not stable numerically. Hence the energy value $-13.31124$ was computed with regularization constant $\epsilon=10^{-6}$, while all the other results were obtained with $\epsilon=10^{-10}$. 
}
\begin{ruledtabular}
\begin{tabular}{rlccccccc}
&&\multicolumn{7}{c}{$M$}
\\[0pt]
\cline{3-9}
\\[-7pt]
Method &$(M_0,M_1,M_2)$	& 4			&5			& 6				& 8				&10 				& 12				& 14  		\\ \\[-7pt] \hline 
\\[-7pt]
\multicolumn{2}{c}{w/o core ($M_0=0$)}  	\\\\[-7pt]
\cline{1-2} \\[-7pt]
TD-RASSCF-S & \hspace{0mm}$(0,3,M-3)$ & $-13.29857^{\sharp}$	& $-13.30037$ 	& $\underline{-13.30039}$		& $\underline{-13.30039}$		& $\underline{-13.30039}$		& $\underline{-13.30039}$		& $\underline{-13.30039}$	  	\\
			&& $(7)$	& $(13)$	& $(19)$		& $(31)$		& $(43)$		& $(55)$		& $(64)$				 		\\
-D & \hspace{0mm}$(0,3,M-3)$	& $-13.29860^{\flat}$	& $-13.30992$ 	& $-13.31749$		& $-13.32618$		& $-13.32717$		& $-13.32730$		& $ -13.32732$	  	\\
			&& $(10)$	& $(43)$ & $(100)$ 	& $(286)$	& $(568)$	& $(946)$	& $(1420)$		 	 		\\
-SD & \hspace{0mm}$(0,3,M-3)$	& $-13.29860^{\flat}$			& $-13.31116$	& $-13.31837$		& $-13.32682$ 		& $-13.32742$ 		& $-13.32751$		& $-13.32753$	  	\\
			& 		& $(16)$ & $(55)$& $(118)$	& $(316)$	& $(610)$	& $(1000)$	& $(1486)$		 	 		\\
-SDT& \hspace{0mm}$(0,3,M-3)$& $$			& $-13.31124$	& $-13.32014$		& $-13.32999$		& $-13.33117$		& $-13.33133$		& $-13.33136$	  	\\
			&& 			& $(91)$	 & $(282)$	& $(1236)$	& $(3326)$	& $(7000)$	& $(12706)$		 	 		\\
MCTDHF & \hspace{0mm}$(0,M,0)$	& $-13.29860^{\flat}$	& $ -13.31127$	& $-13.32016$		& $-13.33009$		& $-13.33133$		& $-13.33151$		& $-13.33154$ 	\\
			&& $(16)$	& $(100)$& $(400)$	& $(3136)$	& $(14400)$	& $(48400)$	& $(132496)$		\\
\\[-7pt]
\hline \\[-7pt]
\multicolumn{2}{c}{w/ core ($M_0=1$)}  	\\\\[-7pt]
\cline{1-2} \\[-7pt]
TD-RASSCF-S & \hspace{0mm}$(1,2,M-3)$ & $-13.29857^{\sharp}$	& $\underline{-13.30037}$& $\underline{-13.30037}$		& $\underline{-13.30037}$		& $\underline{-13.30037}$		& $\underline{-13.30037}$	& $\underline{-13.30037}$  	\\
			&& $(5)$	& $(9)$	& $(13)$		& $(21)$		& $(29)$		& $(37)$	& $(45)$					 	 		\\
-D & \hspace{0mm}$(1,2,M-3)$	& $-13.29860^{\flat}$& $-13.30967$	& $-13.31639$		& $ -13.32383$		&$-13.32431$		& $-13.32439$	& $-13.32440$  	\\
			&& $(5)$	& $(19)$&$(43)$		& $(121)$ 	& $(239)$	 	& $(397)$	& $(595)$			 		\\
-SD & \hspace{0mm}$(1,2,M-3)$	& $-13.29860^{\flat}$			& $-13.31089$	& $-13.31741$		& $-13.32520$		& $-13.32546$ 		& $-13.32550$		& $-13.32551$  	\\
			&& 	$(9)$		& $(27)$	& $(55)$		& $(141)$	& $(267)$	& $(433)$	& $(639)$ 		\\
-SDT& \hspace{0mm}$(1,2,M-3)$	& $$			& $-13.31094$	& $-13.31848$		& $-13.32719$		& $-13.32781$		& $-13.32789$& $-13.32791$	  	\\
			&& 			& $(35)$		& $(91)$		& $(341)$	& $(855)$	& $(1729)$& $(3059)$		 	 		\\
TD-CASSCF & \hspace{0mm}$(1,M-1,0)$	& $-13.29860^{\flat}$& $-13.31094$	& $-13.31848$		& $-13.32722$		& $-13.32786$		& $-13.32795$	& $-13.32796$  	\\
			&& $(9)$	& $(36)$	& $(100)$	& $(441)$	& $(1296)$	& $(3025)$ 	& $(6084)$	\\
\end{tabular}
\end{ruledtabular}
\end{table*}

To illustrate the numerical properties of the TD-RASSCF method, we investigate 1D model atoms defined by the one-body operator
\begin{eqnarray}
h(x,t)=-\frac{1}{2}\frac{d^2}{dx^2}+V(x),
\label{one-body}
\end{eqnarray}
where $V(x)=-Z/\sqrt{x^2+1}$ with $Z=N_{\rm e}=2,4,$ and $6$ for mimicking atomic helium~\cite{Miyagi2014,Hochstuhl2010,Balzer2010,Balzer2010b}, beryllium \cite{Miyagi2013,Miyagi2014,Bonitz2010,Hochstuhl2010b}, and carbon~\cite{Miyagi2014}, respectively. For every atom, the two-body operator,
\begin{eqnarray}
v(x_1,x_2)=\frac{1}{\sqrt{(x_1-x_2)^2+1}},
\end{eqnarray}
is used to describe the electron-electron repulsion. The ground-state wave function was calculated by imaginary-time relaxation~\cite{Kosloff1986} as in Ref.~\cite{Miyagi2013}; a $[-25,25]$ box was discretized by $N_{\rm DVR}=256$ quadrature points associated with Fourier basis functions, and the $\mathcal{Q}$- and $\mathcal{P}$-space orbital equations were regularized with a small constant $\epsilon=10^{-10}$ (see, e.g., Ref.~\cite{Beck2000}). Tables \ref{table_Be} and \ref{table_C} list the ground-state energies of the 1D beryllium and carbon atoms, respectively. The tables also include the number of configurations used to obtain the energies.

We first focus on the results of the 1D beryllium atom in Table \ref{table_Be}. Starting from the HF ground-state energy, $-6.739450$, each method obviously provides smaller energy with increasing $M$ (except TD-RASSCF-S as addressed below). For a given value of $M$, on the other hand, the energy becomes smaller when the number of configurations increases. The MCTDHF method always gives the largest number of configurations and accordingly provides the best energy, which is then followed by the TD-RASSCF-SDT, -SD, -D, and -S methods in this order. In the TD-RASSCF calculations, the  $\mathcal{P}$ space was partitioned as $(M_0,M_1,M_2)=(0,N_{\rm e}/2,M-N_{\rm e}/2)$. When $M=3$, it can be shown by carrying out orbital rotations that the TD-RASSCF-D and MCTDHF methods are equivalent, so that both methods give the same energy value $-6.771296$ as marked by the symbols $\flat$ (see Ref.~\cite{Miyagi2013} for a more detail discussion). Note that the accuracy of the TD-RASSCF-D and -SD methods is comparable. The lack of the single-orbital excitations in the TD-RASSCF-D method seems to be well made up by the orbital optimization. 

Most importantly, the TD-RASSCF-S method, quite differently from the others, shows a peculiar behavior as indicated by the energies underlined in Table~\ref{table_Be}. That is, the TD-RASSCF-S result is converged at $M=4$ to the value $-6.773288$. The special convergence property is stated as a theorem:

\vspace{3mm}
\noindent{\bf Theorem}: {For closed-shell systems, the TD-RASSCF-S method satisfying $M_0+M_1=N_{\rm e}/2$ and $M_1\le M_2$ gives a wave function which is invariant with respect to the value of $M_2$.}

\vspace{3mm}

\noindent The proof is given in Ref.~\cite{Miyagi2014}. The theorem ensures that the TD-RASSCF-S wave function can be fully converged with only $N_{\rm e}/2+1\le M\le N_{\rm e}$ spatial orbitals. Thus without concerns about the convergence with respect to $M$, the TD-RASSCF-S method gives reasonably accurate results for large systems with manageable computational cost. 

The same convergence trend can be observed in the results of the 1D helium atom (not shown) and carbon atom in Table~\ref{table_C}. We briefly consider the results of the 1D carbon atom. In the TD-RASSCF calculations, the  $\mathcal{P}$ space was partitioned as $(M_0,M_1,M_2)=(M_0,N_{\rm e}/2-M_0,M-N_{\rm e}/2)$. The special convergence property of the TD-RASSCF-S method clearly appears in this example as well. When $M=4$, there are some equivalences among the methods for the 1D carbon atom as marked by the symbols $\flat$ and $\sharp$ in Table~\ref{table_C}. The equivalences can be proven by carrying out orbital rotations as discussed above for the 1D beryllium atom~(see Ref.~\cite{Miyagi2013}). The realization of the equivalences by numerical calculations assures the reliability of the numerical results. Finally it should be noticed that, compared to the 1D helium and beryllium atoms, the 1D carbon atom shows more clearly the unfavorable scaling property of the MCTDHF method and emphasizes the efficient reduction of the number of configurations inherent in the other TD-RASSCF methods, especially in the TD-RASSCF-S approach. In Table \ref{table_C}, one can also see the efficiency obtained by introducing a core, $M_0=1$, in terms of reduction in the number of configurations. The use of a core or frozen core will be indispensable for the application of the TD-RASSCF method to very large systems.

\subsection{\label{High-order harmonic spectrum} Laser-induced dynamics}

\begin{figure*}[t]
\begin{center}
\begin{tabular}{c}
\resizebox{170mm}{!}{\includegraphics{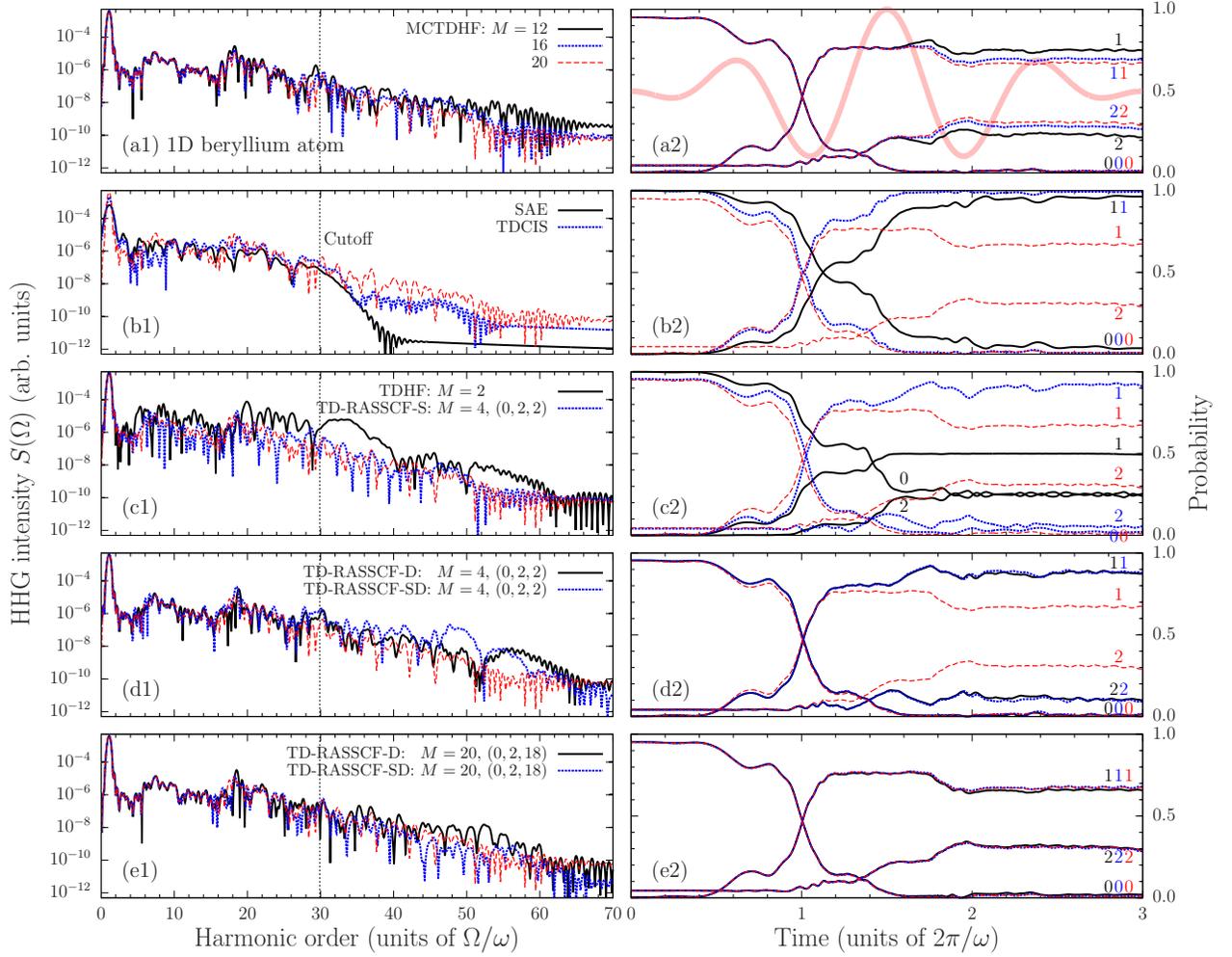}}
\end{tabular}
\caption{
\label{fig_Be_HHG}
(Color online) Left column (a1)--(e1): HHG spectra of the 1D beryllium atom obtained from different methods (see text). Each panel includes the list of methods, and if necessary, also the number of spatial orbitals, $M$, and the partitioning, $(M_0,M_1,M_2)$. Right column (a2)--(e2): Probabilities to find the system in the HF ground state, $\langle P_0\rangle(t)$, single-orbital excited HF states, $\langle P_1\rangle(t)$, and double-orbital excited HF states, $\langle P_2\rangle(t)$ (shortly denoted by `0', `1', and `2', respectively). Left and right panels corresponde to each other, and, for comparison, every panel includes the same dashed (red) lines representing the result of the MCTDHF calculation with $M=20$ spatial orbitals. All the calculations were carried out for the laser pulse specified in Eq.~\eqref{laser}, and the profile is depicted in panel (a2) by the thick (pink) line. For this laser field, the cutoff energy in the HHG spectrum is estimated to be $29.9\omega$ (see text) as shown by the vertical dotted lines in the left column. 
}
\end{center}
\end{figure*}

\begin{figure}[t]
\begin{center}
\begin{tabular}{c}
\resizebox{85mm}{!}{\includegraphics{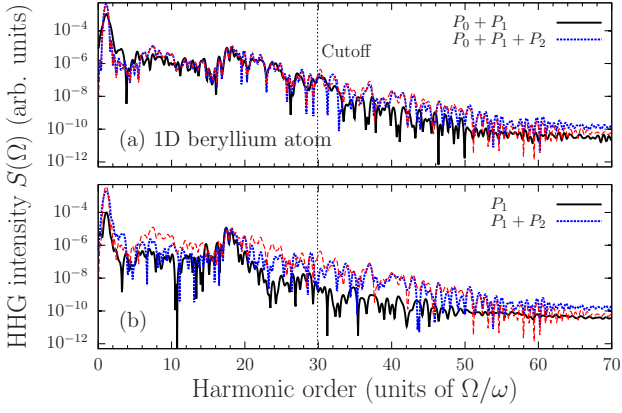}}
\end{tabular}
\caption{
\label{fig_Be_SD}
(Color online) State-resolved analysis of the HHG spectrum of the 1D beryllium atom. For the wave function $|\Psi(t)\rangle$ obtained from the MCTDHF calculation with $M=20$, the state-resolved spectra are computed as follows: (a) the absolute squared of the Fourier transformation of $\langle\Psi(t)|(P_0+P_1)D(P_0+P_1)|\Psi(t)\rangle$ and $\langle\Psi(t)|(P_0+P_1+P_2)D(P_0+P_1+P_2)|\Psi(t)\rangle$, (b) the same for $\langle\Psi(t)|P_1DP_1|\Psi(t)\rangle$ and $\langle\Psi(t)|(P_1+P_2)D(P_1+P_2)|\Psi(t)\rangle$. The reference spectrum of the MCTDHF calculation is shown by dashed (red) line in both panels.
}
\end{center}
\end{figure}

\begin{figure*}[t]
\begin{center}
\begin{tabular}{c}
\resizebox{170mm}{!}{\includegraphics{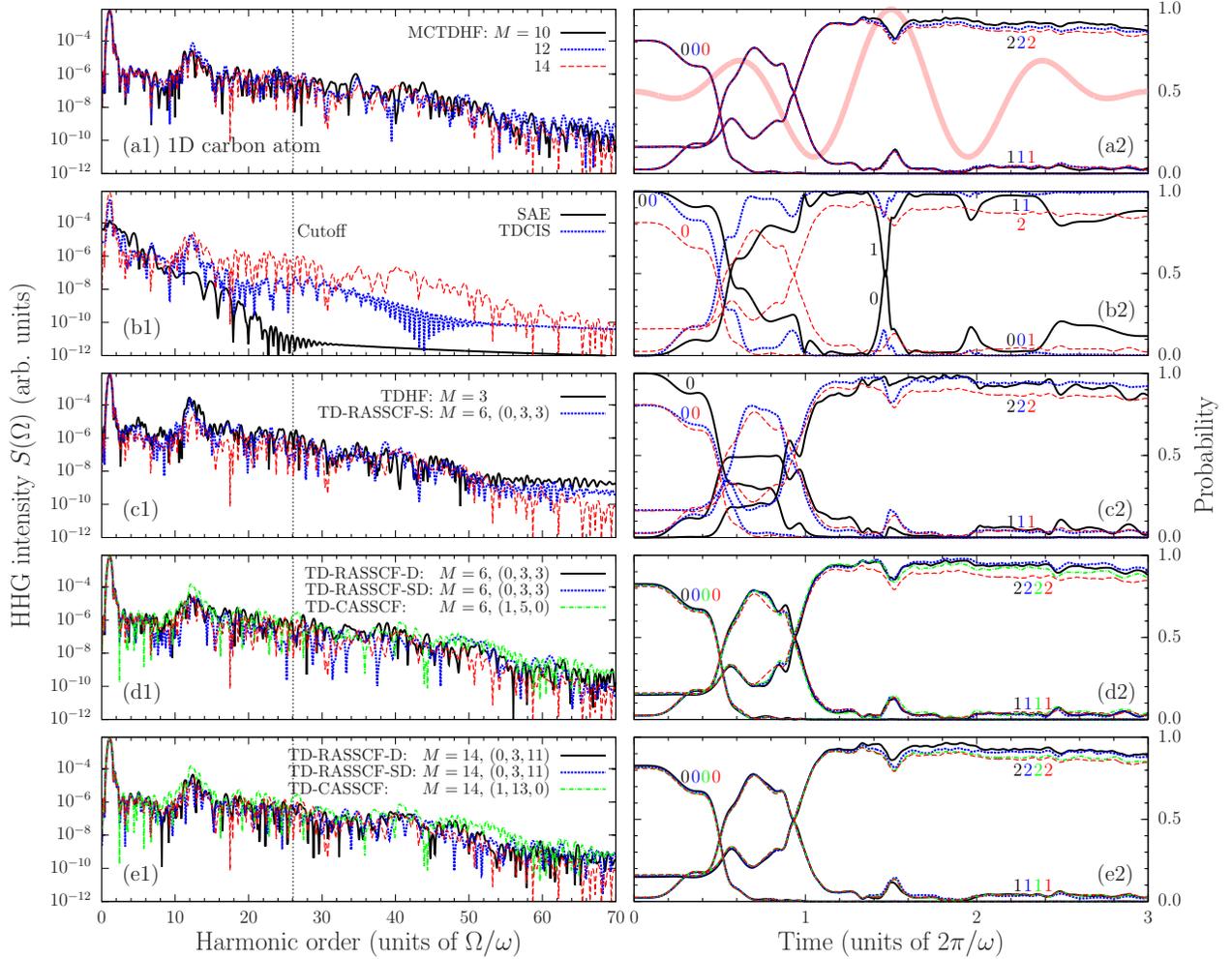}}
\end{tabular}
\caption{
\label{fig_C_HHG}
(Color online) As Fig.~\ref{fig_Be_HHG} but for the 1D carbon atom. The cutoff energy in the HHG spectrum is estimated to be $26.0\omega$.
}
\end{center}
\end{figure*}

\begin{figure}[t]
\begin{center}
\begin{tabular}{c}
\resizebox{85mm}{!}{\includegraphics{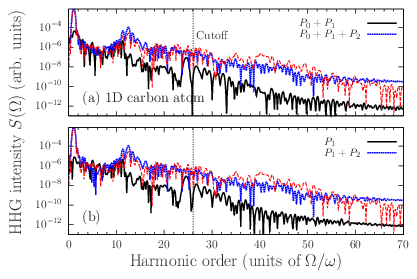}}
\end{tabular}
\caption{
\label{fig_C_SD}
(Color online) As Fig.~\ref{fig_Be_SD} but for the 1D carbon atom. The analysis was carried out for the HHG reference spectrum obtained from the MCTDHF calculation with $M=14$.
}
\end{center}
\end{figure}

We now turn to the discussion of laser-induced dynamics. For the three model atoms introduced above, the real-time propagation was carried out in a large box, $[-300,300]$, discretized by $N_{\rm DVR}=2048$ quadrature points. The electric field is defined as $F(t)\equiv -dA(t)/dt$, where the vector potential is (see, e.g., Ref.~\cite{Y-CHan2010})
\begin{eqnarray}
A(t)=\frac{F_0}{\omega}\sin^2\left(\frac{\pi t}{T}\right)\sin\omega t,
\hspace{2mm} (0\leq t\leq T),
\label{laser}
\end{eqnarray}
with electric field strength, $F_0=0.0755$ ($2.0\times10^{14}$ Wcm$^{-2}$), angular frequency, $\omega=0.0570$ ($800$ nm), and pulse duration $T=331$ ($3$ cycles). Within the dipole approximation, the laser-electron interaction was taken into account in the length gauge, $xF(t)$. However, because of the gauge invariance of the SCF based method (see Appendix~\ref{App-gauge}), the use of the velocity or acceleration gauge causes no change to the dynamics. The other numerical conditions and the definition of the complex absorbing potential (CAP) function~\cite{Kosloff1986cap} are the same as those in Refs.~\cite{Miyagi2013,Miyagi2014}. 

We first focus on the 1D beryllium atom and come back to the other two atoms later. In the TD-RASSCF calculations for the 1D beryllium atom, the partitions are fixed as $(M_0,M_1,M_2)=(0,N_{\rm e}/2,M-N_{\rm e}/2)$ as in the computations of the ground-state energy. The left column of Fig.~\ref{fig_Be_HHG} displays the HHG spectra, $S(\Omega)$, of the 1D beryllium atom computed as the absolute squared of the Fourier transformation of the dipole acceleration (see, e.g., Ref.~\cite{Baggesen2011}), $\langle\Psi(t)|D|\Psi(t)\rangle$, where $D\equiv \sum_{\kappa=1}^{N_{\rm e}}d(x_{\kappa})$ with $d(x_{\kappa})=-dV(x_{\kappa})/dx_{\kappa}$ and $V(x)=-Z/\sqrt{x^2+1}$. Based on the classical model for HHG~\cite{Krause1992,Schafer1993,Corkum1993}, the cutoff energy in the HHG spectrum is estimated to be $3.17U_{\rm p}+I_{\rm p}=29.9\omega$ and indicated by a vertical dotted line. Here $U_{\rm p}=F_0^2/(4\omega^2)=0.439$ is the ponderomotive energy, i.e., the time-averaged energy of a classical free electron quivering in the laser field, and $I_{\rm p}=0.313$ is the first ionization potential based on Koopmans' theorem \cite{Helgaker2000}. To clarify the excitation dynamics during the interaction with the laser, the right column of Fig. \ref{fig_Be_HHG} shows the probabilities to find the system in the HF ground state, single- and double-orbital excited states, $\langle P_{\kappa}\rangle(t)\equiv\langle\Psi(t)|P_{\kappa}|\Psi(t)\rangle$ ($\kappa =0,1$ and $2$) (see Eqs.~\eqref{PT1}--\eqref{PT3} in Appendix~\ref{App-CISD}), computed by using the projection operators:
\begin{eqnarray}
P_0
&=&
|{\rm HF}\rangle \langle{\rm HF}|,
\label{P0} \\
P_1
&=&
\sum_{ia}\sum_{\sigma} |{\rm HF}_{i\sigma}^{a\sigma}\rangle \langle{\rm HF}_{i\sigma}^{a\sigma}|,
\label{P1} \\
P_2
&=&
\frac{1}{4}
\sum_{ijab}\sum_{\sigma\tau}
\Big[
|{\rm HF}_{i\sigma, j\tau}^{a\sigma, b\tau}\rangle\langle {\rm HF}_{i \sigma, j\tau}^{a\sigma, b\tau}|
\nonumber \\
&&
+
(1-\delta_{\sigma}^{\tau})
|{\rm HF}_{i\sigma, j\tau}^{a\tau, b\sigma}\rangle\langle {\rm HF}_{i \sigma, j\tau}^{a\tau, b\sigma}|
\Big],
\label{P2}
\end{eqnarray}
where $|{\rm HF}\rangle$ is the HF ground state, 
$
|{\rm HF}_{i\sigma}^{a\sigma}\rangle
=
{c^{\rm HF}_{a\sigma}}^{\dagger}c^{\rm HF}_{i\sigma}
|{\rm HF}\rangle
$ 
and
$
|{\rm HF}_{i\sigma, j\tau}^{a\sigma, b\tau}\rangle
=
{c^{\rm HF}_{a\sigma}}^{\dagger}{c^{\rm HF}_{b\tau}}^{\dagger}c^{\rm HF}_{j\tau}c^{\rm HF}_{i\sigma}
|{\rm HF}\rangle
$ 
are defined with the annihilation (creation) operator $c^{\rm HF}_{i\sigma}$ (${c^{\rm HF}_{a\sigma}}^{\dagger}$) for the HF occupied (virtual) spin orbitals. The probabilities are plotted after having been divided by the norm squared of the total wave function which is smaller than one due to the CAP. 

The convergence test of the MCTDHF method is demonstrated in Figs.~\ref{fig_Be_HHG} (a1) and (a2). In terms of both the HHG spectrum and the probabilities, one can see the convergence as $M$ increases. Although the complete convergence of the HHG spectrum above the cutoff may require more orbitals, the result with $M=20$ is the most accurate and thus included in the other panels as a reference. Before evaluating the performances of the other methods, we consider Fig.~\ref{fig_Be_SD}, where the state-resolved analysis is carried out for the MCTDHF reference spectrum. Figure \ref{fig_Be_SD} (a) displays the spectrum computed from $\langle\Psi(t)|(P_0+P_1)D(P_0+P_1)|\Psi(t)\rangle$ and $\langle\Psi(t)|(P_0+P_1+P_2)D(P_0+P_1+P_2)|\Psi(t)\rangle$. The insertions of these projection operators select specific electronic processes in terms of field-free time-independent HF orbitals. The detail of the computation is presented in Appendix~\ref{App-CISD}. When $P_0+P_1+P_2$ is inserted in the evaluation of the expectation value of $D$, the original reference spectrum is precisely reproduced over the whole region. One can thus conclude that the triple- and quadruple-orbital excitations in terms of time-independent HF orbitals are not involved in the dynamics governing HHG in the 1D beryllium atom. When $P_0+P_1$ is inserted, on the other hand, the original spectrum is reasonably reproduced below the cutoff, but clearly underestimated beyond. This indicates the important contributions from the double-orbital excitations above the cutoff. Next, to estimate the contribution from the HF ground state, Fig.~\ref{fig_Be_SD} (b) shows the analysis using $P_1$ and $P_1+P_2$. For $P_1+P_2$, the original spectrum is reproduced only above the cutoff. The use of $P_1$ underestimates the spectrum on the whole region. From these observations, one can therefore explain the physic behind the HHG spectrum based on time-independent HF orbitals as follows: Below the cutoff, the recombination processes between the HF ground state and the single-orbital excited states determines the overall shape, and the double-orbital excitations account for the detailed structure. In the region above the cutoff, recombinations among the single- and double-orbital excited states are essential and the contribution from the HF ground state can be negligible. 

Keeping in mind the physical picture revealed by the state-resolved analysis, we proceed with the assessment of the accuracy of each method. Figures~\ref{fig_Be_HHG} (b1) and (b2) show the results of the SAE approximation and the TDCIS method. Note that our SAE approximation as formulated in Appendix~\ref{App-SAE} is a special case of the TDHF or TDCIS method; the nonlocal-exchange interaction between the active electron and the rest of the inactive electrons is taken into account. The SAE approximation reasonably reproduces the HHG spectrum below the cutoff but does not do so above. This is expected from the state-resolved analysis, because the SAE approximation describes only the single-electron recombination process to the HF ground state. The probabilities, $\langle P_0\rangle(t)$ and $\langle P_1\rangle(t)$, obtained from the SAE calculation also clearly differ from the reference ones. On the other hand, the TDCIS method gives improvement to the HHG spectrum above the cutoff and also to $\langle P_0\rangle(t)$ and $\langle P_1\rangle(t)$ by taking into account some multi-orbital effect: coherent contributions from each orbital~\cite{Smirnova2009}, and interchannel interactions among the single-orbital excited states~\cite{Pabst2013}. Due to the lack of multi-orbital excitations, however, the TDCIS method still underestimates the HHG intensity above the cutoff where double-orbital excitations contribute significantly.

Figures \ref{fig_Be_HHG} (c1) and (c2) then show the comparison between the TDHF and TD-RASSCF-S methods. The SCF based methods implicitly take into account any kind of the multi-orbital excitations in terms of time-independent HF orbitals. The TDHF method, however, gives very wrong excitation probabilities and, accordingly, results in an even poorer HHG spectrum than the TDCIS result. On the other hand, the TD-RASSCF-S method shows large improvements. For the computation of the HHG spectrum especially above the cutoff, the TD-RASSCF-S method is obviously more accurate than the TDHF and TDCIS methods. By the explicit inclusion of the single-orbital excitations in terms of the TD orbitals, the TD-RASSCF-S method succeeds in describing the double-orbital excitations in terms of time-independent HF orbitals. Note that, by the theorem stated in Sec.~\ref{Ground-state wave function} and Ref.~\cite{Miyagi2014}, the TD-RASSCF-S result is fully converged with $M=N_{\rm e}$. As shown in Ref.~\cite{Miyagi2014}, the converged TD-RASSCF-S wave function with $M=N_{\rm e}$ is more accurate than the TDCIS wave function in the sense of the TD variational principle. The numerical results in Fig.~\ref{fig_Be_HHG} reflect this property of the theory. Also note that, whereas the TDCIS method is gauge dependent, the SCF based method is gauge independent as shown in Appendix~\ref{App-gauge}. This is another advantage of the TD-RASSCF method compared to the TDCIS approach or the TDCI methods with truncation in excitation level.

Next Figs.~\ref{fig_Be_HHG} (d1) and (d2) compare the performances of the TD-RASSCF-D and -SD methods for $M=4$. Despite the lack of explicit inclusion of single-orbital excitations in the TD-RASSCF-D method, the accuracy of the TD-RASSCF-D and -SD methods is almost comparable and they are slightly more accurate than the TD-RASSCF-S method. The same computations were carried out with $M=20$ and the results in Figs.~\ref{fig_Be_HHG} (e1) and (e2) show the expected variational improvements. The HHG spectra and the excitation probabilities are in excellent agreement with the MCTDHF reference values. The TD-RASSCF-D computation with $M=20$ is more expensive than the TD-RASSCF-S and -D computations with $M=4$ but more economical than the MCTDHF approach with $M=20$.

To assess the accuracy of the TD-RASSCF methods in more detail, we next consider the results for the 1D helium and carbon atoms. The MCTDHF, TDHF, TD-RASSCF-S, and TDCIS methods and the SAE approximation were used for the 1D helium atom to compute the HHG spectra and excitation probabilities, and the results were compared with the exact solution to the TD Schr\"odinger equation (TDSE). Due to the large ionization potential, $I_{\rm p}=0.750$, and small polarizability of the 1D helium atom, many-electron effects are of minor importance, and the HHG spectra and the excitation probabilities obtained from all the methods are in reasonable agreement (see Ref.~\cite{Miyagi2014} where the acceleration dipoles and HHG spectra obtained from the MCTDHF, TD-RASSCF-S, TDCIS methods are shown). On the contrary, due to the small ionization potential, $I_{\rm p}=0.093$, and large polarizability, the 1D carbon atom requires a more accurate treatment of the many-electron effects and serves as a critical test case to assess the accuracy of the methods. Figure~\ref{fig_C_HHG} displays the HHG spectra and excitation probabilities computed for the 1D carbon atom (see also Ref.~\cite{Miyagi2014}). The most accurate result obtained from the MCTDHF method with $M=14$ is used as a reference. As shown in Fig.~\ref{fig_C_HHG}, the failure of the SAE approximation and the TDCIS and TDHF methods is apparent even in the region below the cutoff $3.17U_{\rm p}+I_{\rm p}=26.0\omega$. On the other hand, the accurate performance of the TD-RASSCF-S, -D, and -SD methods is emphasized. The TD-CASSCF method with $M_0=1$ also performs accurately as shown in Figs.~\ref{fig_C_HHG} (d1)--(e2). To unveil the dynamics governing the HHG process a state-resolved analysis was carried out for the MCTDHF reference HHG spectrum in the same manner as for the 1D beryllium atom discussed above. Figure~\ref{fig_C_SD} (a) shows that the reference spectrum is roughly reproduced when $P_0+P_1+P_2$ is used as the projector but not when $P_0+P_1$ is used. Thus over the whole region, double- and higher-order orbital excitations play an essential role. Figure~\ref{fig_C_SD} (b) shows the state-resolved analysis with $P_1$ and $P_1+P_2$. The absence of large changes between panels (a) and (b) indicates the negligible role of the HF ground state in the HHG process even below the cutoff. This fact is understandable from the rapid depletion of $\langle P_0\rangle(t)$ and the accompanying rises of $\langle P_1\rangle(t)$ and $\langle P_2\rangle(t)$ in Fig.~\ref{fig_C_HHG} (a2). Since the SAE approximation and the TDCIS method dismiss the effect of multi-orbital excitations and overestimate the contributions from the HF ground state and the single-orbital excited states, their break-down is therefore natural. The TD-RASSCF-S, -D, and -SD methods and the TD-CASSCF method still succeed in taking into account multi-orbital excitations accurately.

\subsection{\label{Accuracy} Relations among the methods based on the time-dependent variational principle}

To finalize this section, we specify what `accurate' means in connection to the relation among the methods considered in this paper. Based on the TD variational principle, the more variational parameters a method includes, the more `accurate' the method is. We define the wording `accurate' in this context. This is a way (if not the only) to give meaning to the word `accurate' in TD problems. Hence, for example, TD-RASSCF-S is more accurate than TDHF. For a given number of spatial orbitals, $M$, and fixed partitions, $(M_0,M_1,M_2)$, TD-RASSCF-SD is more accurate than TD-RASSCF-S, TD-RASSCF-SDT is more accurate than TD-RASSCF-SD, and so on so forth, and ultimately TD-CASSCF, or MCTDHF (when $M_0=0$), is the most accurate among the series. It is also true that TD-RASSCF-SD is more accurate than TD-RASSCF-D. The relation between TD-RASSCF-S and TDCIS is nontrivial at first glance. For closed shell systems, however, the converged TD-RASSCF-S wave function with $M=N_{\rm e}$ is more accurate than the TDCIS wave function as theoretically demonstrated in Ref.~\cite{Miyagi2014}. Meanwhile the relation between TDHF and TDCIS remains unclear. While the HHG spectra of the 1D beryllium atom in Figs.~\ref{fig_Be_HHG} (b1) and (c1) indicate that TDCIS is more accurate than TDHF, the HHG spectra of the 1D carbon atom in Figs.~\ref{fig_C_HHG} (b1) and (c1) may give an opposite impression. Their relative accuracy could change for different target systems or for different laser parameters. The relation between TD-RASSCF-S and -D is likewise unclear. Based on the computation of the ground-state energy in Sec.~\ref{Ground-state wave function} and the laser-induced dynamics in Sec.~\ref{High-order harmonic spectrum}, TD-RASSCF-D is, however, practically more accurate than TD-RASSCF-S. In view of the favorable scaling properties, we therefore concluded that the TD-RASSCF-S and -D methods will be efficient tools for studying the TD many-electron problem with an accuracy higher than the TDHF and TDCIS methods.

\section{\label{Conclusion} Conclusions}

As a generalization of the MCTDHF and TD-CASSCF methods, we have developed the TD-RASSCF method. The key idea is the use of both RAS and SCF schemes, by which the number of the CI-expansion coefficients and orbitals can be reduced, and large systems which are impossible to study by the MCTDHF method can be addressed. Following the formulation of the TD-RASSCF-D method in Ref.~\cite{Miyagi2013}, we presented a more generalized framework, i.e., a series of methods, TD-RASSCF-S, -SD, and -SDT. The numerical cost analysis and test calculations for the 1D helium, beryllium, and carbon atoms showed the TD-RASSCF-S and -D methods were computationally feasible for large systems and more accurate than the TDHF and TDCIS approaches. 

In addition to the methodological progress, we reported a state-resolved analysis of the HHG spectrum for the 1D beryllium atom based on the field-free time-independent HF orbitals. For the accurate MCTDHF calculation with $M=20$, the analysis clarified the significant contribution from the recombination processes between the HF ground state and single-orbital excited states in the region below the cutoff, but among the single- and double-orbital excited states beyond. This observation rationalized why the SAE approximation and the TDCIS method failed in the computation of the HHG spectrum in the region above the cutoff but succeeded below. On the other hand, the TD-RASSCF-S and -D methods succeeded in describing the multi-orbital excitations accurately. The state-resolved analysis was carried out also for the 1D carbon atom, and it was shown that multi-orbital excitations were more essential for the HHG processes even below the cutoff in this model than for 1D beryllium. The break-down of the SAE approximation and the TDCIS method was thus emphasized while the TD-RASSCF-S and -D methods remained accurate.

In summary, the TD-RASSCF-S and -D methods will be useful numerical tools for analyzing the nonperturbative many-electron dynamics. In particular, for investigating the laser-induced dynamics, the gauge independence is another advantage. The TD-RASSCF-S and -D methods could open a new perspective in intense laser research fields by elucidating the role of electron correlation in large atoms and molecules.


\begin{acknowledgments}
This work was supported by the ERC-StG (Project No. 277767-TDMET), and the VKR center of excellence, QUSCOPE.
\end{acknowledgments}

\appendix

\section{\label{App-SCALE} Scaling property}

In this Appendix, the number of operations at each time step for solving the equations of motion of the TD-RASSCF approach is estimated. Consider an $N_{\rm e}$-electron system ($N_{\rm e}$ is even). In the MCTDHF method, the dimension of the Fock space is expressed as a function of $N_{\rm e}$ with a parameter $M(\ge N_{\rm e}/2)$, denoting the number of $\mathcal{P}$-space orbitals. Since in the MCTDHF method there is no restriction on the orbital excitations, the dimension is given as the number of all possible ways to distribute the electrons among the orbitals,
\begin{eqnarray}
{\rm MCTDHF:}
\hspace{3mm}
\dim\mathcal{V}_{\rm MHF}(N_{\rm e};M)
=
\left(
\begin{array}{c}
M\\
N_{\rm e}/2\\
\end{array}
\right)^2.
\end{eqnarray}
In the TD-RASSCF method with partitions, $(M_0,M_1,M_2)=(0,N_{\rm e}/2,M-N_{\rm e}/2)$, the dimension of the Fock space is expressed as follows:
\begin{eqnarray}
&&\hspace{-20mm}{\rm TD\mbox{-}RASSCF\mbox{-}S:}
\hspace{-0mm}
\nonumber \\
&&
\hspace{-15mm}
\dim\mathcal{V}_{\rm S}(N_{\rm e};M)=
1
+
\dim\mathcal{V}_1(N_{\rm e};M),
\\
&&\hspace{-20mm}{\rm TD\mbox{-}RASSCF\mbox{-}D:}
\hspace{0mm}
\nonumber \\
&&
\hspace{-15mm}
\dim\mathcal{V}_{\rm D}(N_{\rm e};M)=
1
+
\dim\mathcal{V}_2(N_{\rm e};M),
\\
&&\hspace{-20mm}{\rm TD\mbox{-}RASSCF\mbox{-}SD:}
\hspace{0mm}
\nonumber \\
&&
\hspace{-15mm}
\dim\mathcal{V}_{\rm SD}(N_{\rm e};M)=
1
+
\dim\mathcal{V}_1(N_{\rm e};M)
\nonumber \\
&&
\hspace{17mm}
+
\dim\mathcal{V}_2(N_{\rm e};M),
\\
&&\hspace{-20mm}{\rm TD\mbox{-}RASSCF\mbox{-}SDT:}
\nonumber \\
&&
\hspace{-15mm}
\dim\mathcal{V}_{\rm SDT}(N_{\rm e};M)=
1
+
\dim\mathcal{V}_1(N_{\rm e};M)
\nonumber \\
&&
\hspace{-10mm}
+\dim\mathcal{V}_2(N_{\rm e};M)
+
\dim\mathcal{V}_3(N_{\rm e};M),
\end{eqnarray}
For clarity, the notation of the Fock space, $\mathcal{V}$, used in the main text is replaced by the method specific notation. Based on Eq.~\eqref{direct_sum_K}, the dimension of each subspace, $\mathcal{V}_n(N_{\rm e};M)$ ($n=1,2,$ and $3$), is calculated as
\begin{align}
\dim\mathcal{V}_1(N_{\rm e};M)=&
2(N_{\rm e}/2)(M-N_{\rm e}/2),
\\
\dim\mathcal{V}_2(N_{\rm e};M)=&
\Big[(N_{\rm e}/2)(M-N_{\rm e}/2)\Big]^2
\nonumber \\
&\hspace{-0mm}
+
2
\left(
\begin{array}{c}
N_{\rm e}/2\\
2\\
\end{array}
\right)
\left(
\begin{array}{c}
M-N_{\rm e}/2\\
2\\
\end{array}
\right),
\\
\dim\mathcal{V}_3(N_{\rm e};M)=&
2
\left(
\begin{array}{c}
N_{\rm e}/2\\
3\\
\end{array}
\right)
\left(
\begin{array}{c}
M-N_{\rm e}/2\\
3\\
\end{array}
\right)
\nonumber \\
&\hspace{-20mm}
+2(N_{\rm e}/2)(M-N_{\rm e}/2)
\left(
\begin{array}{c}
N_{\rm e}/2\\
2\\
\end{array}
\right)
\left(
\begin{array}{c}
M-N_{\rm e}/2\\
2\\
\end{array}
\right).
\end{align}

Using the expressions defined above, for 1D model systems like the ones discussed in Sec.~\ref{Numerical performance}, the number of operations at each time step for solving the equations of motion is roughly estimated as follows (see the discussion in Sec.~\ref{Numerical scaling}):
\begin{widetext}
\begin{align}
{\rm TDHF:}&
&f_{\rm HF}(N_{\rm e};N_{\rm DVR})&=
(N_{\rm e}/2)^4N_{\rm DVR}^2,
\label{dimV_HF}
\\
{\rm TD\mbox{-}RASSCF\mbox{-}S:}&
&f_{\rm S}(N_{\rm e};M,N_{\rm DVR})&=
2M^4
\Big( N_{\rm DVR}^2
+\dim\mathcal{V}_{\rm S}(N_{\rm e};M)\Big)
+M^6\dim\mathcal{V}_1(N_{\rm e};M),
\label{dimV_S}
\\
{\rm \mbox{-}D:}&
&f_{\rm D}(N_{\rm e};M,N_{\rm DVR})&=
2M^4
\Big( N_{\rm DVR}^2
+\dim\mathcal{V}_{\rm D}(N_{\rm e};M)\Big),
\label{dimV_D}
\\
{\rm \mbox{-}SD:}&
&f_{\rm SD}(N_{\rm e};M,N_{\rm DVR})&=
2M^4
\Big( N_{\rm DVR}^2
+\dim\mathcal{V}_{\rm SD}(N_{\rm e};M)\Big)
+M^6\dim\mathcal{V}_2(N_{\rm e};M),
\label{dimV_SD}
\\
{\rm \mbox{-}SDT:}&
&f_{\rm SDT}(N_{\rm e};M,N_{\rm DVR})&=
2M^4
\Big( N_{\rm DVR}^2
+\dim\mathcal{V}_{\rm SDT}(N_{\rm e};M)\Big)
+M^6\dim\mathcal{V}_3(N_{\rm e};M),
\label{dimV_SDT}
\\
{\rm MCTDHF:}&
&f_{\rm MHF}(N_{\rm e};M,N_{\rm DVR})&=
2M^4
\Big(N_{\rm DVR}^2
+\dim\mathcal{V}_{\rm MHF}(N_{\rm e};M)\Big).
\label{dimV_FCI}
\end{align}
\end{widetext}
In Sec.~\ref{Numerical scaling}, Fig.~\ref{fig_scale} displays the behaviors of the functions~\eqref{dimV_HF}--\eqref{dimV_FCI}. In the plot, the parameters are set to be $N_{\rm DVR}=2048$ and $M=N_{\rm e}$. Note that, for closed-shell systems, the TD-RASSCF-S wave function is fully converged with $M=N_{\rm e}$ (see the theorem in Sec.~\ref{Ground-state wave function} and Ref.~\cite{Miyagi2014}).

\section{\label{App-CISD} State-resolved analysis}

In this Appendix, some details are shown for the state-resolved analysis based on field-free time-independent HF orbitals. When the wave function is multiplied by the projection operator $P_0+P_1$ [Eqs.~\eqref{P0} and \eqref{P1}], the expectation value of the acceleration dipole reads (see also Ref.~\cite{Rohringer2006}),
\begin{eqnarray}
&&
\hspace{-10mm}
\langle\Psi(t)|(P_0+P_1)D(P_0+P_1)|\Psi(t)\rangle
\nonumber \\
&&
\hspace{-6mm}
=2\sum_{i=1}^{N_{\rm e}/2}
\Bigg\{
2{\rm Re}\Big[
\langle{\rm HF}|\Psi(t)\rangle
\langle\chi_i(t)|d|\phi_i^{\rm HF}\rangle\Big]
\nonumber \\
&&
\hspace{6mm}
-\sum_{j=1}^{N_{\rm e}/2}d^j_i\langle\chi_i(t)|\chi_j(t)\rangle
+\langle\chi_i(t)|d|\chi_i(t)\rangle
\Bigg\},
\label{CISD1-2}
\end{eqnarray} 
where $d^j_i=\langle\phi^{\rm HF}_j|d|\phi^{\rm HF}_i\rangle$ with the HF occupied orbitals $|\phi^{\rm HF}_i\rangle$ ($i=1,\cdots,N_{\rm e}/2$). The one-electron wave packet is introduced as
\begin{eqnarray}
\chi_{i}(x,t)
\equiv
\sum_{n=1}^M \tilde{\phi}_n(x,t) \langle {\rm HF}|{c_{i\sigma}^{\rm HF}}^{\dagger} c_{n\sigma}|\Psi(t)\rangle
\hspace{2mm}
(\sigma=\uparrow \:{\rm or}\: \downarrow),
\nonumber \\
\label{CIS-extraction}
\end{eqnarray}
where ${c_{i\sigma}^{\rm HF}}^{\dagger}$ is an electron creation operator in the spin orbital $|\phi_{i\sigma}^{\rm HF}\rangle\equiv |\phi_i^{\rm HF}\rangle\otimes|\sigma\rangle$, and $\tilde{\phi}_n(x,t)$ is the $\mathcal{P}$-space orbital defined by orthogonalization to occupied HF orbitals as
\begin{eqnarray}
|\tilde{\phi}_n(t)\rangle\equiv 
\left(
1-\sum_{i=1}^{N_{\rm e}/2}
|\phi^{\rm HF}_i\rangle\langle\phi^{\rm HF}_i|
\right)
|\phi_n(t)\rangle.
\end{eqnarray}
The one-electron integrals in Eq.~\eqref{CISD1-2} should be performed for the spatial coordinate, $x$.

When the wave function is multiplied by $P_0+P_1+P_2$ [Eqs.~\eqref{P0}, \eqref{P1}, and \eqref{P2}], the expectation value of the acceleration dipole reads
\begin{eqnarray}
&&
\hspace{-0mm}
\langle\Psi(t)|(P_0+P_1+P_2)D(P_0+P_1+P_2)|\Psi(t)\rangle
\nonumber \\
&&
\hspace{0mm}
=
\langle\Psi(t)|(P_0+P_1)D(P_0+P_1)|\Psi(t)\rangle
\nonumber \\
&&
\hspace{0mm}
+
\sum_{i,j=1}^{N_{\rm e}/2}
\sum_{\sigma,\tau=\uparrow,\downarrow}
\Bigg\{
2{\rm Re}\Big[
\langle\lambda_{i\sigma,j\tau}|D|\phi_{i\sigma}^{\rm HF}\chi_{j\tau}(t)\rangle
\Big]                      
\nonumber \\
&&
\hspace{0mm}
-
2\sum_{k=1}^{N_{\rm e}/2}
d^k_j\langle\lambda_{i\sigma,j\tau}(t)|\lambda_{i\sigma,k\tau}(t)\rangle 
+
\langle\lambda_{i\sigma,j\tau}(t)|D|\lambda_{i\sigma,j\tau}(t)\rangle
\Bigg\},
\nonumber \\
\label{CISD3}
\end{eqnarray}
where the two-electron wave packets, $|\lambda_{i\sigma,j\tau}(t)\rangle$ and $|\phi_{i\sigma}^{\rm HF}\chi_{j\tau}(t)\rangle$, are introduced such that the spin-spatial representations are defined by
\begin{eqnarray}
&&
\langle z_1\;z_2|\lambda_{i\sigma,j\tau}(t)\rangle
\nonumber \\
&&
\hspace{3mm}
\equiv
\frac{1}{2}
\sum_{n,m=1}^M \sum_{\sigma',\tau'=\uparrow,\downarrow}
\Big\|\tilde{\phi}_{n}(x_1,t)\sigma'(s_1)\: \tilde{\phi}_{m}(x_2,t)\tau'(s_2)\Big\|
\nonumber \\
&&
\hspace{25mm}
\times
\langle {\rm HF}|{c_{i\sigma}^{\rm HF}}^{\dagger} {c_{j\tau}^{\rm HF}}^{\dagger} c_{m\tau'}c_{n\sigma'}|\Psi(t)\rangle,
\label{CISD-extraction}
\\
&&\langle z_1\;z_2|\phi_{i\sigma}^{\rm HF}\chi_{j\tau}(t)\rangle
\equiv
\Big\|\phi_i^{\rm HF}(x_1)\sigma(s_1)\: \chi_j(x_2,t)\tau(s_2)\Big\|,
\nonumber \\
\end{eqnarray}
with $\big\|\cdots\big\|$ being the normalized Slater determinant, $z_1\equiv (x_1,s_1)$ and $z_2\equiv (x_2,s_2)$ being the spin-spatial coordinates for which the two-electron integrals in Eq.~\eqref{CISD3} are performed. Finally note that the values of $\langle\Psi(t)|P_1DP_1|\Psi(t)\rangle$ and $\langle\Psi(t)|(P_1+P_2)D(P_1+P_2)|\Psi(t)\rangle$ for computing the spectra in Fig.~\ref{fig_Be_SD} (b) and Fig.~\ref{fig_C_SD} (b) are obtained from Eqs.~\eqref{CISD1-2} and \eqref{CISD3}, respectively, by removing the term $4{\rm Re}\big[\langle{\rm HF}|\Psi(t)\rangle\sum_{i}\langle\chi_i(t)|d|\phi_i^{\rm HF}\rangle\big]$. 

The one- and two-electron wave packets defined by Eqs.~\eqref{CIS-extraction} and \eqref{CISD-extraction}, respectively, are important not only for the state-resolved analysis of HHG but also for computing the probabilities to find the system in the HF ground state, or single- or double-orbital excited HF states. These probabilities are given as follows:
\begin{eqnarray}
\langle\Psi(t)|P_0|\Psi(t)\rangle&=&
|\langle{\rm HF}|\Psi(t)\rangle|^2,
\label{PT1}
\\
\langle\Psi(t)|P_1|\Psi(t)\rangle&=&
2\sum_{i=1}^{N_{\rm e}/2} \langle\chi_{i}(t)|\chi_{i}(t)\rangle,
\label{PT2}
\\
\langle\Psi(t)|P_2|\Psi(t)\rangle&=&
\frac{1}{2}\sum_{i,j=1}^{N_{\rm e}/2}\sum_{\sigma,\tau=\uparrow,\downarrow}
\langle \lambda_{i\sigma,j\tau}(t)|\lambda_{i\sigma,j\tau}(t)\rangle.
\nonumber \\
\label{PT3}
\end{eqnarray}

\section{\label{App-SAE} Single-active-electron approximation based on the HF approximation}

\begin{figure}[t]
\begin{center}
\begin{tabular}{c}
\resizebox{65mm}{!}{\includegraphics{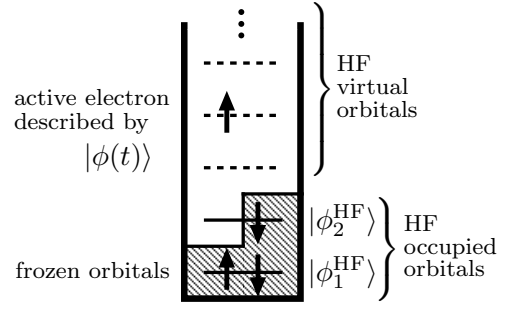}}
\end{tabular}
\caption{
\label{fig_SAE}
Illustration of the SAE approximation considered  in this paper. The HF occupied orbitals, $\big\{|\phi_i^{\rm HF}\rangle\big\}_{i=1}^{N_{\rm e}/2}$, are numbered in ascending order from the lowest energy. There is one active electron described by a wave packet $|\phi(t)\rangle$ whose initial condition is $|\phi(0)\rangle=|\phi_{N_{\rm e}/2}^{\rm HF}\rangle$. The rest of the electrons always occupy the frozen orbitals in shaded area. This illustration shows an example for a four-electron system. 
}
\end{center}
\end{figure}

The SAE approximation in this paper is a special case of the TDHF or TDCIS method. The solution of the HF equations gives a set of HF occupied orbitals, which is denoted by $\big\{|\phi_i^{\rm HF}\rangle\big\}_{i=1}^{N_{\rm e}/2}$. By freezing all the occupied spin orbitals except one of the highest-energy spin orbitals, $|\phi_{N_{\rm e}/2}^{\rm HF}\rangle\otimes|\sigma\rangle$ ($\sigma=\uparrow$ or $\downarrow$), we consider the time evolution of a one-electron wave packet $|\phi(t)\rangle$ with the initial condition $|\phi(t=0)\rangle\equiv |\phi_{N_{\rm e}/2}^{\rm HF}\rangle$ as illustrated in Fig.~\ref{fig_SAE}. The equation of motion is derived from the TD variational principle and given as
\begin{eqnarray}
iQ|\dot{\phi}(t)\rangle
=
Qh(t)|\phi(t)\rangle
+
Q(\mathcal{J}(t)-\mathcal{K}(t))|\phi(t)\rangle,
\label{App-SAE-1}
\end{eqnarray} 
where $Q=1-\sum_{i=1}^{N_{\rm e}/2-1}|\phi_i^{\rm HF}\rangle\langle\phi_i^{\rm HF}|-|\phi(t)\rangle\langle\phi(t)|$, and the Coulomb and exchange potential operators, respectively, are defined by
\begin{eqnarray}
&&\mathcal{J}(t)=2\sum_{i=1}^{N_{\rm e}/2-1}\langle\phi_i^{\rm HF}|v|\phi_i^{\rm HF}\rangle
+\langle\phi_{N_{\rm e}/2}^{\rm HF}|v|\phi_{N_{\rm e}/2}^{\rm HF}\rangle,
\label{App-SAE-2} \\
&&
\mathcal{K}(t)|\phi(t)\rangle=\sum_{i=1}^{N_{\rm e}/2-1}\langle\phi_i^{\rm HF}|v|\phi(t)\rangle |\phi_i^{\rm HF}\rangle.
\label{App-SAE-3}
\end{eqnarray} 

In the SAE approximation usually a time-independent and local effective potential function is designed using fitting parameters such that the ground-state one-electron wave function formed in the potential imitates the property of the highest-energy occupied orbital of the target system (see, e.g., Ref.~\cite{Abu-samha2010}). However, the present SAE approximation uses the TD nonlocal potential operators, $\mathcal{J}(t)-\mathcal{K}(t)$, and no fitting parameter is included (see Ref.~\cite{Hochstuhl2012} for a related SAE approximation). Note that, the exchange potential operator, $\mathcal{K}(t)$, takes into account the interaction between the laser-driven active electron and the rest of the electrons in the frozen orbitals.

\section{\label{App-gauge} Gauge independence of the TD-RASSCF method}

The TD-RASSCF method is gauge independent. To show this important feature, consider an $N_{\rm e}$-electron system interacting with light fields. The TDSE reads $i\partial_t|\Psi^{\rm G}(t)\rangle=H^{\rm G}(t)|\Psi^{\rm G}(t)\rangle$ (G$=$L, V, and A), where L, V, and A, respectively, mean the length, velocity, and acceleration gauges within the dipole approximation. The exact solution to the TDSE is gauge independent, i.e., the solutions in different gauges are related by unitary transformations. Between the velocity and length gauges, for instance, the wave functions are related by $|\Psi^{\rm V}(t)\rangle=U(t)|\Psi^{\rm L}(t)\rangle$, where
\begin{eqnarray}
\hspace{-5mm}
U(t)=
\exp\left[
-i\sum_{\kappa=1}^{N_{\rm e}}{\bm A}(t)\cdot{\bm r}_{\kappa}
+\frac{iN_{\rm e}}{2}\int^t{\bm A}^2(t')dt'
\right].
\end{eqnarray}
This is not always the case when the wave function is expressed approximately.

Next consider the TD-RASSCF wave function,
\begin{eqnarray}
|\Psi_{\rm SCF}^{\rm G}(t)\rangle
=
\sum_{I\in\mathcal{V}}
C^{\rm G}_I(t)|\Phi_I^{\rm G}(t)\rangle,
\label{Gauge1}
\end{eqnarray} 
where each Slater determinant $|\Phi_I^{\rm G}(t)\rangle$ is composed of spin orbitals $|\phi^{\rm G}_i(t)\rangle\otimes|\sigma\rangle$ ($i=1,\cdots,M$). The sets of the CI-expansion coefficients $\big\{C^{\rm G}_I(t)\big\}_{I\in\mathcal{V}}$ and the spatial orbitals $\big\{|\phi^{\rm G}_i(t)\rangle\big\}_{i=1}^M$, respectively, obey the amplitude and orbital equations for the gauge G. Since Eq.~\eqref{Gauge1} is not the exact solution to the TDSE, it is \textit{a priori} unclear whether the TD-RASSCF wave functions in different gauges are exactly unitarily related. To see their relations, we consider as an example the velocity and length gauges. In this case, the set of equations of motion [Eqs.~\eqref{amp_orb}, \eqref{Q_orbital_eq1}, and \eqref{P_orb_eqs5}] are given for the one-body Hamiltonians, $h^{\rm V}(t)=-\frac{1}{2}\nabla^2+V({\bm r})-i{\bm A}(t)\cdot\nabla$ and $h^{\rm L}(t)=-\frac{1}{2}\nabla^2+V({\bm r})+{\bm F}(t)\cdot{\bm r}$. It can easily be checked that, by defining $C_I^{\rm V}(t)=C_I^{\rm L}(t)$ and
\begin{eqnarray}
&&|\phi_i^{\rm V}(t)\rangle
\equiv
\exp\left(
-i{\bm A}(t)\cdot{\bm r}
+\frac{i}{2}\int^t{\bm A}^2(t')dt'
\right)
|\phi_i^{\rm L}(t)\rangle,
\nonumber \\
&&\hspace{50mm}
(i=1,\cdots,M),
\label{gauge2}
\end{eqnarray}
the sets of equations of motion in both gauges are unitarily transformed into each other. It follows immediately from Eq.~\eqref{gauge2} that $|\Phi_I^{\rm V}(t)\rangle=U(t)|\Phi_I^{\rm L}(t)\rangle$, thus
\begin{eqnarray}
|\Psi_{\rm SCF}^{\rm V}(t)\rangle
&=&
\sum_{I\in\mathcal{V}}
C^{\rm V}_I(t)|\Phi_I^{\rm V}(t)\rangle
\nonumber
\\
&=&
\sum_{I\in\mathcal{V}}
C^{\rm V}_I(t)U(t)|\Phi_I^{\rm L}(t)\rangle
\nonumber
\\
&=&
U(t)|\Psi_{\rm SCF}^{\rm L}(t)\rangle.
\end{eqnarray}
The unitary relations between the other pairs of gauges are also obviously true. The TD-RASSCF method is therefore gauge independent. It should be emphasized that this is concluded for every TD-RASSCF method, i.e., TDHF, MCTDHF, TD-CASSCF, and TD-RASSCF-S, -D, -SD, -SDT, $\cdots$. It is not essential for the discussion but, if an arbitrary phase factor comes into Eq.~\eqref{gauge2}, it should be compensated by including the inverse phase factor in the definition of the CI-expansion coefficients. This is another degree of freedom inherent in the TD-RASSCF method as discussed in Refs.~\cite{Miyagi2013,Lubich2008}.

Note that the TDCIS method is, on the other hand, gauge dependent due to the use of time-independent HF orbitals as basis functions. This unfavorable fact can be seen as follows. First define the TDCIS wave function in the length gauge as
\begin{eqnarray}
|\Psi_{\rm CIS}^{\rm L}(t)\rangle
=
\alpha_0(t)|{\rm HF}\rangle
+
\sum_{ia}\alpha_i^a(t)|{\rm HF}_i^a\rangle.
\end{eqnarray} 
The TDCIS wave function in the velocity gauge $|\Psi_{\rm CIS}^{\rm V}(t)\rangle$ is expressed in the same TDCIS ansatz, but they are not related by a unitary transformation. In fact, $U(t)|\Psi_{\rm CIS}^{\rm L}(t)\rangle$ is expressed by the full-CI expansion, not the CIS;
\begin{eqnarray}
&&U(t)|\Psi_{\rm CIS}^{\rm L}(t)\rangle
=
U(t)\left(
\alpha_0(t)|{\rm HF}\rangle
+
\sum_{ia}\alpha_i^a(t)|{\rm HF}_i^a\rangle
\right)
\nonumber
\\
&&
\hspace{5mm}
=
{\rm TD}\;{\rm full \mathchar`- CI}\;{\rm ansatz}
\ne
{\rm TDCIS}\;{\rm ansatz},
\end{eqnarray} 
because each HF orbital is transformed as
\begin{eqnarray}
&&\exp\left(
-i{\bm A}(t)\cdot{\bm r}
+\frac{i}{2}\int^t{\bm A}^2(t')dt'
\right)
|\phi_q^{\rm HF}\rangle
\nonumber
\\
&&\hspace{20mm}
=
\sum_{p}u_{pq}(t)|\phi_p^{\rm HF}\rangle,
\end{eqnarray} 
where $u_{pq}(t)=
\langle\phi_p^{\rm HF}|
\exp\left(
-i{\bm A}(t)\cdot{\bm r}
+\frac{i}{2}\int^t{\bm A}^2(t')dt'
\right)
|\phi_q^{\rm HF}\rangle$. Hence the TDCIS wave function depends on the choice of the gauge. More generally, except the TD full-CI expansion method, the use of time-independent basis functions leads to gauge dependent results.


\end{document}